*Analysis of the "Big Bang" and the Resulting Outward Cosmic Expansion:*
*Hubble - Einstein Cosmology vs. The Universal Exponential Decay*

by

*Roger Ellman*

Abstract

There is general agreement that the universe began with an outward "explosion" of matter and energy at a "singularity" followed by its on-going expansion -- the "Big Bang". This paper analyses the mechanics of that beginning and two alternative theories related to it:

- The Hubble - Einstein theory that that beginning created space itself, expanding and carrying the universe's matter and energy with it such that the velocity, $v$, of recession of any distant astral object from us is directly proportional to its distance, $v = H_0 \cdot d$, where $H_0$ is the "Hubble Constant", and

- The Universal Decay theory that the length, $[L]$, dimensional aspect of all quantities in the universe [e.g. distance $[L]$, speed $[L/T]$, gravitation constant, $G$, $[L^3/M \cdot T^2]$ etc.] is exponentially decaying while the material universe is expanding outward within passive, static "space".

The Hubble - Einstein theory has been accepted by consensus for many decades.

The Universal Decay theory was first propounded in "*The Origin and Its Meaning*"[1] in 1996 and has since been validated by the Pioneer 10 and 11 "anomalous acceleration" as well as by the theory's success in accounting for "dark matter" and "dark energy"[2]. The same centrally-directed, distance-independent acceleration of $(8.74 \pm 1.33) \times 10^{-8}$ $cm/s^2$ that is the Pioneer "anomalous acceleration" supplies the "additional gravitation" that "dark matter" is sought to supply and is part of the on-going contraction and decay of all length, $[L]$, dimensions.

The mass of the universe is calculated and the universe's "Schwarzschild Radius" evaluated. The velocities and distances of cosmic objects in general are calculated and plotted from the end of the "inflation" to the present. It develops that there is a theoretical limit on how far back into the past can be observed regardless of the quality of our instrumentation.

Roger Ellman, The-Origin Foundation, Inc.
 320 Gemma Circle, Santa Rosa, CA 95404, USA
 RogerEllman@The-Origin.org
 http://www.The-Origin.org

*Analysis of the "Big Bang" and the Resulting Outward Cosmic Expansion:
Hubble - Einstein Cosmology vs. The Universal Exponential Decay*
*by*
*Roger Ellman*

PART I -- THE THEORIES

*1 - The Nature of Space*

There is general agreement that the universe began with and results from an essentially instantaneous appearance and outward "explosion" of the matter and energy of the universe at a "singularity", and the resulting on-going expansion -- the "Big Bang".

*a. Hubble - Einstein Theory*

The Hubble - Einstein theory, thoroughly and extensively elaborated in numerous books and scientific papers, is that the result of that beginning was the creation of space, itself, and that it is space, itself, that is expanding, and in the process carrying the universe's matter and energy along with it -- an expanding universe such that the velocity, $v$, of recession of any distant astral object from us, the observers, is directly proportional to the object's distance from us, $v = H_0 \cdot d$, where $H_0$ is the "Hubble Constant", the value of which has been not well determined beyond being in the range of $50 - 100$ $km/sec$ $per$ $megaparsec$, but has been recently reported per analysis of a Hubble Space Telescope survey as $72$ $km/sec$ $per$ $megaparsec$ [3].

In spite of the long term acceptance of the Hubble - Einstein cosmological concept there are fundamental questions about it that are unanswered. The concept is a direct result of Einstein's General Theory of Relativity for which space, itself, is some kind of "substance" [not Einstein's terminology] capable of expanding and capable of being "curved" by the effect on it of gravitating masses in it. That concept leaves the problem, "... relative to what" ? If space is expanding then the expansion must be relative to some static, non-expanding reference. If space is curved than the curvature must be relative to some flat, uncurved reference. One cannot have relativity without relativity. Any change or effect must be relative to a previous unchanged reference or previous unaffected state. Otherwise the change or effect would be undetectable.

So, what do we call that "static, flat, uncurved reference"? It is space itself; and it is, and it must be, the framework that expansion of the universe is relative to. And flat, uncurved rectilinear space is and must be, the framework that curved motion due to gravitation is relative to. And that space must have always existed unoccupied [and, therefore actually "nothing"] until the "Big Bang" introduced matter and energy into it.

Furthermore, were space, itself, to be expanding as in the Hubble - Einstein theory, then it would be expanding everywhere including the expansion of the space containing and within all of our measurement standards and instrumentation [and selves]. But, an expanding ruler used to measure an expanding universe would report only a static state, not an expansion. The expansion would not be detectable by us if it were space, itself, that is expanding. Since we detect the expansion, then it must be, and is, the objects within space that are moving away from each other [away from the "Big Bang" location]. And, therefore, space itself is passive and static.

*b. The Universal Decay Theory*

The Universal Decay theory that the universe is exponentially decaying while its content is expanding outward within static "space" was first propounded in "*The Origin and Its Meaning*"[1] in 1996 and has since been validated by the Pioneer 10 and 11 "anomalous acceleration" as well as by the theory's success in accounting for "dark matter" and "dark energy"[2]. The same centrally- directed, distance-independent acceleration of $(8.74 \pm 1.33) \times 10^{-8}$ $cm/s2$ that is the Pioneer "anomalous acceleration" supplies the "additional gravitation" that "dark matter" is sought to supply and is part of the on-going contraction and decay of all length, $[L]$, dimensions. The details, mechanisms, and parameters are all thoroughly developed in the references. For the present purposes the pertinent aspects are as follows.

The Universal Decay theory is that the length, $[L]$, dimensional aspect [with all dimensions expressed in the fundamental dimensions of mechanics, $[L]$, $[M]$, and $[T]$] of all quantities in the universe [e.g. distance $[L]$, speed $[L/T]$, gravitation constant, $G$, $[L^3/M \cdot T^2]$, Planck constant, $h$, $[M \cdot L^2/T]$, etc.] is exponentially decaying with time in the form:

*(1)* $\quad L(t) = L(0) \cdot \varepsilon^{-t/\tau}$

and the value of the time constant, $\tau$, [predicted in "*The Origin and Its Meaning*"[1] from theoretical calculations and validated by the Pioneer 10 and 11 "anomalous acceleration"[2]] is

*(2)* $\quad \tau = 3.57532 \cdot 10^{17} \text{ sec} \quad [\approx 11.3373 \cdot 10^9 \text{ years}]$

Because it is the length, $[L]$, dimensional aspect of all quantities that is decaying, all of those quantities remain consistent with each other through the laws of physics that connect them. The very long time constant compared to a human life time would make our observation of the decay difficult. And, in any case, because all of our measuring instrumentation and we ourselves are undergoing that same decay, we cannot directly measure or observe it.

One would expect that exactly the corresponding argument to the above presented last criticism of the hypothesis that space, itself, is expanding -- specifically that the Universal Decay should be undetectable because the measuring equipment is also decaying -- would prevent ever observing the decay. However, in the case of the Pioneer 10 and 11 satellites and the case of galactic rotation curves, the decay has been detected because it forces orbital / path behavior that would not be present if there were no decay, and that orbital / path behavior can be and has been observed.

The Universal Decay is the principal cause of redshifts. There must be some Doppler content in redshifts because the astral sources do have velocities away from us, the observers, but it can readily be shown that the Doppler effect accounts for only about 10% or less of the total redshift. The Universal Decay produces the redshifts because when we observe light from distant astral sources we are observing light emitted long ago, which means that it was emitted less decayed than our local contemporary light. Less decay of the length, $[L]$, dimensional aspect of all quantities in that light means that its speed and wavelength are greater than the spectrally corresponding light from sources local to us. Being less decayed it appears redshifted to us because we compare it to our local, decayed-to-date light and spectra. The decay is of the length, $[L]$, dimensional aspect, which affects the wavelength. There is no decay of the period, $[T]$, nor its inverse, the frequency.

There is no violation of invariance involved. At any instant of time the speed of the light emitted by any and every source anywhere and everywhere in the universe is the same, that is the as-of-that-time current decayed value, because all decay started at the same time, the instant of the "Big Bang", and all decay is with the same decay constant, $\tau$. However, once emitted the light continues propagating onward at the speed at which it was emitted. The decay is in the generation, is in the source, not the propagation. That is the case because the emitted light carries within it its own propagation-determining permeability and dielectric constant, $\mu$ and $\varepsilon$. How could it be otherwise since light propagating outward into unoccupied space, into pure nothing, would certainly find no $\mu$ and $\varepsilon$ there: in nothing ?

As a consequence, therefore, direct measurement of the speed and Planck's constant $[c$ and $h]$ in ancient light is another way to observe the decay because it applies our decayed equipment to light that is much less decayed.

### c. *Comparison of the Two Theories*

A quick calculation makes clear that, in a universe in which the redshifts are dominantly due to the Universal Decay with only a minor [Hubble - Einstein] Doppler content, the age of the universe must be considerably greater than the current estimates, that are based on Hubble - Einstein reasoning, of about `13.7 billion years`. The detection of redshifts of `z ≈ 10` has recently been reported[5]. Those would require an age of the universe [a time period to encompass sufficient multiples of the decay time constant $\tau$] of about `30 billion years`.

In Hubble - Einstein reasoning, the recent observations at those increasingly greater redshifts are nearing the point of leaving little or no time, between the instant of the very beginning and the first appearance of the most ancient galaxies, little or no time for the "Dark Ages" and initial galaxy formation to have taken place -- a process estimated only a few years ago to have required and have taken `2 - 3 billion years` and now reduced by default subtraction [the Hubble - Einstein estimated age of the universe less the Hubble - Einstein estimated age of high redshift light observed] to as little as a `few 100 million years`. And, observation of still greater redshifts with the resulting further compression of the [Hubble - Einstein] time available for the "dark ages" may be shortly forthcoming.

Further in favor of the Universal Decay theory, is that exponential decay is found essentially everywhere in physics, in nature. It would almost seem to be a requirement of a universe coming into existence with a sudden "bang". The Hubble - Einstein cosmological concept appears to be in severe and increasing trouble. It has always suffered from the absence of competition with an alternative "Big Bang" cosmology [its only significant competition was the now defunct steady state universe theory]. In addition, the useful Occam's Razor [the simplest explanation is most likely the correct one] is certainly against Hubble - Einstein.

## 2 - The Topology of the Universe

### a. Hubble - Einstein

In the Hubble - Einstein conception the universe that arose from the "Big Bang" has no "center" and no "edge"; rather, space -- the universe -- is a topologically "closed space" with nothing else beyond its "spatial limits". And, therefore, it is finite.

The theory's space having come into being from the original singularity and having continuously expanded thereafter, it is difficult to justify the concept that the location of the original singularity is nowhere, and in particular that it is not somewhere within the expanding space of the universe that arose from it. Likewise, it is difficult to justify the concept of the expanding universe of space having no edge, no boundary. Because it is expanding, at any moment the universe's "closed space" encloses a smaller "closed space" that it was a moment ago. Some distinction between the two is necessary else there would be no expansion. The enclosed smaller "closed space" must have a boundary or "spatial limits" that distinguish it from the total enclosing larger "closed space" with its larger "spatial limits". If the time interval between the two approaches zero the boundaries of the enclosed become the boundaries of the enclosing.

Or, is the evidence of the expansion the increasing separation distances of the objects within the "closed space"? If so there is no difference between that and the objects traveling outward from their original source through a passive, static space that is not a "closed space".

Thus, while a boundaryless "closed space" can be theoretically conceived of, such a space that is non-static, that is expanding, is impossible as a material reality. Doing away with a boundaryless "closed space" universe with nothing outside of it returns the discussion to the problem: If the universe is finite then what is outside of it, and should not that be called the universe, and must not the universe then be infinite ? But a material infinity is impossible -- how are the two reconciled ? See "The Origin and Its Meaning", below.

Theorist mathematicians like to use analogies to justify their space with no edge. They cite the surface of a sphere as a two-dimensional space having no edge, no boundary, as it exists in the three-dimensional space of the sphere whose surface it is, and they then ask that that example be extrapolated to a three-dimensional universe in a four-dimensional space. But the sphere surface is in three dimensions and has two boundaries: [1] the boundary between the surface of the sphere and everything outside of the sphere and [2] the boundary between the surface of the sphere and everything inside the sphere. [Some theorists like to cite the Moebius Strip, an example of quibbling with definitions, not an example of different space.]

### b. *The Origin and Its Meaning*

The Origin and Its Meaning conception of the cosmic topology is that space is a three-dimensional Euclidean metric which is nothing until something occupies it. It is now not nothing because part of it is occupied by the matter and energy of the universe. The metric extends infinitely in all directions, but only a finite portion is occupied by the universe. The unoccupied portion is nothing, only a metric.

The exponential decay of the universe is just that; it is not a decay of the metric in which the universe resides. The decay is relative to the metric.

The universe is finite:

- Finite now in that it has been expanding from a point source outward in all directions at a finite speed for a finite amount of time.

- Finite forever, in that the exponential decay of the upper limit on that finite speed, the speed of light, $c$, makes it be $c(t)$ as equation *(3)*,

*(3)*  $c(t) = c(0) \cdot \varepsilon^{-t/\tau}$

which produces a finite ultimate result as shown in equation *(4)*.

*(4)*  $\int_0^\infty c(t) \cdot dt = c(0) \cdot \tau$

## *3 - The Origin of the "Big Bang"*

There is general agreement that the universe had to come into being from a preceding "nothing" at a "singularity".

The reason for the universe's arising from nothing is that the alternative is the prior existence of some not-nothing, something, which then must have its own existence accounted for. The only ultimate "beginning" is truly "nothing".

The reason for requiring an original singularity is that otherwise there would be an infinite rate of change in getting from nothing to something. [Yet, getting an entire universe into existence through the portal of a dimensionless singularity is another concern in the overall problem].

[Incidentally, the entire reasoning above demonstrates that we cannot do without metaphysics and that in a sense metaphysics is superior to physics in that metaphysics can address problems not accessible by physics and metaphysics can lay a basis for broad aspects of physics. Physics itself must depend on reproducible verifiable observations; but the validating of hypotheses based on and resulting from such physics data, especially in the absence of competing alternative hypotheses during the development and validation phase, can be subject to error.]

### a. *20th Century Physics*

The 20th Century Physics theory for the origin of the "Big Bang" does not really exist in that physics insists that its purview extends only to that which can be observed and measured, and we have not and probably can not observe and measure the origin of the "Big Bang".

However, "unofficial" ideas on the subject rely on thinking that is a result of the combination of quantum mechanics and Heisenberg uncertainty where the latter is considered to be actual, real, probabilistic uncertainty, not Heisenberg's original point that measurement is uncertain because the process of measuring changes that which is measured.

The 20th Century Physics thinking is that the "nothing" from which the universe arose was not quite "nothing" because of quantum variations and uncertainty. Rather, the primal "nothing" is deemed a "quantum foam" with particles continuously flashing into and out of existence. That is nevertheless deemed to be "nothing" because on the average the "quantum foam" is neutral or zero. It is theorized to have produced the universe from a fortuitous flashing into existence.

However, a "quantum foam" even though on the average neutral or zero is not quite the same thing as pure "nothing". The existence of the "quantum foam" still needs to be accounted for. 20th Century Physics' contention in that regard is that the primal "quantum foam" naturally always existed by its nature and the laws of quantum mechanics and uncertainty -- that it was as "nothing" as could possibly ever be. How an entire universe of such particles arose in an instant at a "singularity" and so as to proceed on its expansion, remains a problem.

### b. The Origin and Its Meaning

The Origin and Its Meaning conception of the origin of the "Big Bang" is that before the "Big Bang" there was nothing -- absolute nothing with no characteristics nor content -- but that even such a nothing cannot have an infinite duration. Consequently the nothing divided into something and an equal but opposite un-something, thus maintaining conservation while changing sufficiently to interrupt the original nothing's otherwise infinite duration. The universe is the combined result of both of the mutually off-setting "halves" acting jointly.

The event was of extremely low probability, but its probability was not zero -- it was possible. Therefore, by the process of the original nothing's duration approaching the infinite, the original nothing provided opportunity approaching the infinite for the extremely low probability of its change event. Consequently it ultimately occurred, which occurrence was an unavoidable ultimate necessity and which occurrence we call the "Big Bang".

Resolution of the problem of an entire universe appearing and expanding outward through the portal of a dimensionless singularity is fully developed and resolved in *The Origin and Its Meaning*[1] as is, also, the problem of why the mutually off-setting "halves" did not promptly cancel each other by re-combining.

## PART II -- THE EXPANSION OF THE UNIVERSE FROM THE "BIG BANG" OUTWARD

On the basis of the foregoing, the Hubble - Einstein conception of the universe is not valid and the Universal Decay conception of the universe is valid and that validity has already been demonstrated by experiment and observation. The following analysis of the expansion of the universe outward following the first instant of the "Big Bang" is in terms of the Universal Decay conception of the universe as set forth in summary above. The analysis is of the mechanics of the travel of matter outward from its "Big Bang" source [some of it ultimately being we the observers] and of the mechanics of the travel of light from such material sources wherever they are at the time that the light that we later observe is emitted. Results of the analysis are that there is a limit to how far back into the past we are theoretically able to observe [quite regardless of the quality of our observation instrumentation] and that the age of the universe is at least about 30 billion years.

### 1 - The Travel of Matter and Light

The first step is to develop formulations that describe the travel of the two different traveling entities, light and matter, at various times in the past from at the beginning to the present.

The <u>travel of matter</u> originated at the location of the "Big Bang" singularity and was initially radially outward from that location.  While mass cannot travel at light speed the initial speed of the "Big Bang" product particles was sufficiently near the then [initial un-decayed] light speed so as to be taken as such as is developed below.  Two effects then proceeded to slow the outward velocities:  the decay of the speed of light [the upper limit on particle velocity] and the gravitational slowing [the centrally directed gravitational acceleration, caused by the total mass, decelerating the outward velocities].

The treatment here is of the estimated "average" or "typical" cosmic body [e.g. galaxy], treated as that from its initial form as myriad fundamental particles at the instant of the "Big Bang" -- the particles ultimately destined to form that particular "typical" body, through its form as we know it now.  [While not of concern in the present analysis, once the outward travel began the particles experienced local gravitational effects in addition to the overall general slowing -- effects that deflected paths from being purely radially outward and that lead to "clumping" and the formation of structure in the universe.]

The <u>travel of light</u> originated from the above traveling matter, at its various locations and times throughout the universe from the first instant on.  It was radially outward from wherever its source was at the time of emission.  Its speed was the speed of light at the decayed value for the time after the "Big Bang" that the light was emitted.

### a. *The Travel of Light Outward From Astral Sources*

Astral / cosmic source light emitted long ago was emitted at a higher "light speed" than our local contemporary light and continues to travel at that faster speed forever as explained above under the sub-heading of "Universal Decay".  On the other hand, the matter originating with the "Big Bang" cannot have traveled at light speed [because its mass would then be infinite] other than nearly so initially before being slowed by gravitation.  Therefore, all cosmic source light has been traveling at greater speeds than the cosmic bodies that are home for observers of the light.

Consequently, the most ancient light that it would be theoretically possible for us to observe would be light from a cosmic source that exited the "Big Bang" in the diametrically opposite direction to that of the planetary home [or its components before they became the home planet] of we, the observers.  That way, the ancient light has to travel a maximally greater distance from its location where and when emitted to our location where and when we observe it than did our planetary home have to travel from its location when the light was emitted to its location when we observe the light.  In other words, ancient light is light that has been traveling a long time and, therefore, has traveled a great distance.  The home of we, the observers cannot travel so fast and must, therefore, have a "head start" of distance to be able to arrive at the meeting place of light and observer at the same moment as the faster light.  The largest "head start" is the handicapping of the light by placing its source diametrically opposite the location of the observers.

Standard International [SI] units are used; however the great range of magnitudes of the quantities considered calls for their being expressed sometimes in alternative astronomical units:  time in *Gyrs = Years·10⁹* rather than *seconds* and distances in *"our" G-Lt-Yrs = 10⁹ × [Light Years at our contemporary speed of light]* rather than *meters*.  [Note: G-Lt-Yrs is always "our" G-Lt-Yrs.]  Those are obtained by the following factors.

(5)  $k_{time} = 60 \cdot 60 \cdot 24 \cdot 365 \cdot 10^9$   *seconds per giga year*   $[sec/_{Gyr}]$

$k_{dist} = k_{time} \cdot$ *["Our" Light Speed]*   *meters per giga light Year*

$= k_{time} \cdot [2.997,924,58 \cdot 10^8]$   $[meters/_{G-Lt-Yr}]$

For notes concerning precision see References [4].

For the present the age of the universe is taken to be unknown so that *Age* is a variable. Then based on equation *(3)*, the original speed of light, *c(0)*, at the instant of the "Big Bang", just before the first moment of the Universal Decay, is obtained as in equation *(6)*, below.

*(6)*    $c(t) = c(0) \cdot \varepsilon^{-t/\tau}$   meters/sec      [light decay per equation *(3)*]

       $c(Age) \equiv 2.997,924,58 \cdot 10^8$ meters/sec      ["our" c, now]

       $c(Age) = c(0) \cdot \varepsilon^{-Age/\tau}$      [set t = Age in equation*(3)*]

       $c(0) = c(Age) \cdot \varepsilon^{+Age/\tau}$      [solve for c(0)]

            $= 2.997,924,58 \cdot 10^8 \cdot \varepsilon^{Age/\tau}$ meters/sec

Then, the speed of light at any arbitrary time, *t*, after the "Big Bang" for any arbitrary age of the universe, *Age*, is as follows.

*(7)*    $c(t, Age) = c(0) \cdot \varepsilon^{-t/\tau} = [2.997,924,58 \cdot 10^8 \cdot \varepsilon^{Age/\tau}] \cdot \varepsilon^{-t/\tau}$   meters/sec

### b. *The Travel of Cosmic Bodies Outward From the Origin of the "Big Bang"*

To determine the travel of cosmic bodies outward from the "Big Bang" one needs to know the initial velocities and the manner in which they subsequently were reduced by gravitation and other effects. The initial radially outward velocities were so close to the then speed of light as to be that speed for the practical precision here being used. That determination develops as follows.

#### (1) *The Initial Radially Outward Velocities*

The universe has existed for billions of years and is still expanding. Therefore, the initial velocity / energy of the "Big Bang" product particles must have been near, if not at or greater than, the escape velocity / energy. The escape velocity / energy for any one particle of the initial "Big Bang" universe is calculated as follows. [The calculation is done non-relativistically here and consequently produces apparent velocities much greater than that of light. They represent velocities nearly at light speed with greatly increased mass.]

<u>Gravitational escape velocity</u> is that velocity the kinetic energy of which just equals in magnitude the potential energy of position in the gravitational field for which the escape velocity is being determined. The non-relativistic escape velocity of a particle develops as follows.

*(8)*    Kinetic Energy = Potential Energy = Force × Distance

                         = Gravitational Attraction × Particle Center to Universe Center Distance

Using:   $v_{esc} \equiv$ escape velocity
          $m_p \equiv$ mass of the particle
          $m_U \equiv$ mass of the Universe [after the initial, essentially instantaneous, mutual annihilations]
          $d_0 \equiv$ distance [from the center of mass of the particle to the center of mass of the universe]
          $G \equiv$ gravitation constant [un-decayed original value at the time of the "Big Bang"]

Then:

$$\tfrac{1}{2} \cdot m_p \cdot v_{esc}^2 = G \cdot \left[\frac{m_p \cdot m_U}{d_0^2}\right] \times d_0$$

$$v_{esc} = \left[\frac{2G \cdot m_U}{d_0}\right]^{\tfrac{1}{2}}$$

For that formulation the needed data are: the gravitation constant, $G$, the mass of the universe, $m_U$, and the separation distance, $d_0$. Estimating the <u>Mass of the Universe</u>, $m_U$, proceeds by estimating the average mass density, $\rho$, and the volume. The universe mass is then the product of the two. The <u>mass density of the universe</u>, $\rho$, develops as follows.

Astronomical analyses treat a "critical density" of the universe, $\rho_c$, which is the particular value of the average density that is on the boundary separating the case of an open (expanding forever) versus closed (eventually gravitationally recontracting) universe. The critical density relates to the escape velocity presented in equation *(8)*, above. The development begins with equating kinetic and potential energy in the form of the next to last line of that equation as in equation *(9)*, below.

*(9)*
$$\tfrac{1}{2} \cdot m_p \cdot v^2 \;=\; G \cdot \left[ \frac{m_p \cdot m_U}{d} \right]$$

The "Hubble Law" states that the velocity of an astral object is proportional to its distance. That law, where $H_0$ is the "Hubble Constant", is

*(10)*  $v = H_0 \cdot d$

The total mass inside a sphere of radius $d$ is

*(11)*  $M = [\text{Volume}] \cdot [\text{density}] = [\tfrac{4}{3} \cdot \pi \cdot d^3] \cdot [\rho]$

Substituting in equation *(9)* for $v$ with equation *(10)* and for $m_U$ with equation *(11)* the result is as follows.

*(12)*
$$\tfrac{1}{2} \cdot m_p \cdot [H_0 \cdot d]^2 \;=\; G \cdot \left[ \frac{m_p \cdot [[\tfrac{4}{3} \cdot \pi \cdot d^3] \cdot \rho]}{d} \right]$$

$$\rho \;=\; \frac{3 \cdot H_0^2}{8 \cdot \pi \cdot G} \qquad \text{[Simplifying and solving for } \rho\text{]}$$

That formulation is intended to give the average density of a portion of the universe of volume $\tfrac{4}{3} \cdot \pi \cdot d^3$ such that the mass is on the boundary between escape from that volume and ultimate recapture. It would also, then, be the critical average density, $\rho_c$, for the overall universe, except for the following problem.

The very concept of the "Hubble Constant" is only valid in terms of the Hubble - Einstein theory that it is space itself that is expanding. It is that which would, if valid, justify the concept of one number, a "universal constant", representing the ratio of distance to velocity. The analogy given for the Hubble - Einstein concept of $H_0$ is that of the blowing up of a balloon or the rising of a loaf of bread in both of which examples the separation velocity of two locations within is proportional to their separation distance.

However, the "Hubble Constant" and concept are not valid, as already presented. The *form* of equation *(12)* is valid and correct, but the constant, $H_0$, must be replaced with a valid number, the correct ratio of distance to velocity for the object the escape of which is being considered, and that number is not a constant but, rather, depends on the particular circumstances.

Even in Hubble - Einstein terms, the "Hubble Constant", $H_0$, would better be referred to as the "Hubble Parameter". Not even the first digit of its numerical value is securely determined and its value has been taken to be over a range of from less than $H_0 = 50$ to nearly $H_0 = 100$ for various calculations and estimates by various researchers.

Further, in the "Hubble Law", $v = H_0 \cdot d$, the distance $d$ is the distance of the astral object from the *observer*. The correct distance for the form of equation *(12)*, that is the distance as in Universal Decay terms not Hubble - Einstein terms, is the distance *outward from the origin* of the "Big Bang". In other words, the "law" is that the object's outward velocity from the origin of the "Big Bang" must be, and must have been, relatively faster if its distance outward, the time-integral of that velocity, is greater, which is obvious. The "Hubble Law" is correct to that extent, but only to that.

Of course the Hubble - Einstein cosmology involves even greater error in attributing redshifts solely to the Doppler Effect of the astral object's velocity rather than the dominant cause, the Universal Decay. That means that determinations of the distance of astral objects by taking their outward velocity from the redshift as a purely Doppler effect, an incorrect velocity, and multiplying it by $H_0$ , an invalid number and concept, can produce only distances in error.

Because for lesser distances from now back into the past [perhaps to `4 or 5 Gyrs` ago] Hubble - Einstein redshift calculations of distance deviate relatively less from the correct Universal Decay calculations a look at the results given by equation `(12)` may nevertheless be somewhat helpful in estimating universe average mass density. Depending on the value of $H_0$ used in equation `(12)`, various values for the mass density $\rho$ result, for example:

```
Value of ρ with the now favored H₀ = 72 km/sec/megaparsec:
    ρ = 9.8·10⁻²⁷ kg/m3
Value of ρ with the past favored H₀ = 49 km/sec/megaparsec:
    ρ = 4.5·10⁻²⁷ kg/m3
```

On the other hand estimates of $\rho$, rather than theoretical calculations as just above, have been made by estimating the mass of a typical galaxy, that done by estimating the number of stars in a galaxy and multiplying by the estimated average star mass and considering the galaxy's rotational dynamics; then counting the number of galaxies in a volume of space, the process performed for increasingly larger volumes. That procedure has produced a universe mass density estimate of:

```
Value of ρ from estimating star mass densities:
    ρ ≈ 10⁻²⁷ kg/m3
```

Having, then, estimates ranging from about `1` to `10` times $10^{-27}$, a reasonable value to use for the mass density of the universe would be the average, about:

*(13)*     $\rho_U \approx 5 \cdot 10^{-27}$ kg/m3

Next the volume of the universe is needed so as to obtain the universe's mass as the product of the mass density and the volume. The <u>volume of the universe</u> develops as follows.

The particles of matter of the universe cannot have commenced their travel outward from the origin of the "Big Bang" at one same speed; rather their initial speeds must have been over a range of speeds, which would have produced a wide distribution in space as their travel developed. While the preceding analysis has developed an *average* mass density for the universe, $\rho$, the actual density must vary substantially even on the scale of large volumes. Therefore, to address the issue of to what volume the average mass density is to be applied requires addressing the issue of the distribution of the initial velocities of the matter emerging from the "Big Bang" because that velocity distribution is the cause of the spatial distribution of astral objects.

The analysis further on below related to Figure 2c, *First Phase of The Expansion of The Universe -- Velocities for Age = 30 Gyrs Case*, shows that the limits on the range of initial energies of those emerging particles set that range to energies of about `0 to 3,000 × [the escape energy]`. Those limits are the obvious lower limit of zero and an upper limit of energy so great that the matter fails to slow to non-relativistic speed ever. However, that is a very large range. Even to only `1,000` is quite large. The range used here for sample cases will be from `1 to 1,000 × [the escape energy]`. We cannot know the exact distribution of those energies; however, there are known energy distributions of other natural phenomena that can be a guide.

The energy distributions considered are that of Planck Black Body Radiation and of the Maxwell - Boltzman treatment of the kinetic theory of gases. Replacing the case-specific constants [ $\pi$, $h$, $c$, $k$, $2$, and parameter $T$ ] with summary case-neutral constants the form of those distributions is as in equation *(14)*, and they appear as in Figures 1a and 1b, below.

```
(14)  Where:  F is relative energy [multiple Factor of escape energy].
              n(F) is the number of particles of energy multiple F.
              p(F) is the probability of interval [F+ΔF], ΔF→0.
```

Planck:
$$n(F) = \frac{K1 \cdot F^5}{\varepsilon^{K2 \cdot F} - 1}$$

Maxwell-Boltzman:
$$p(F) = \frac{K3 \cdot F^{\frac{1}{2}}}{\varepsilon^{K4 \cdot F}}$$

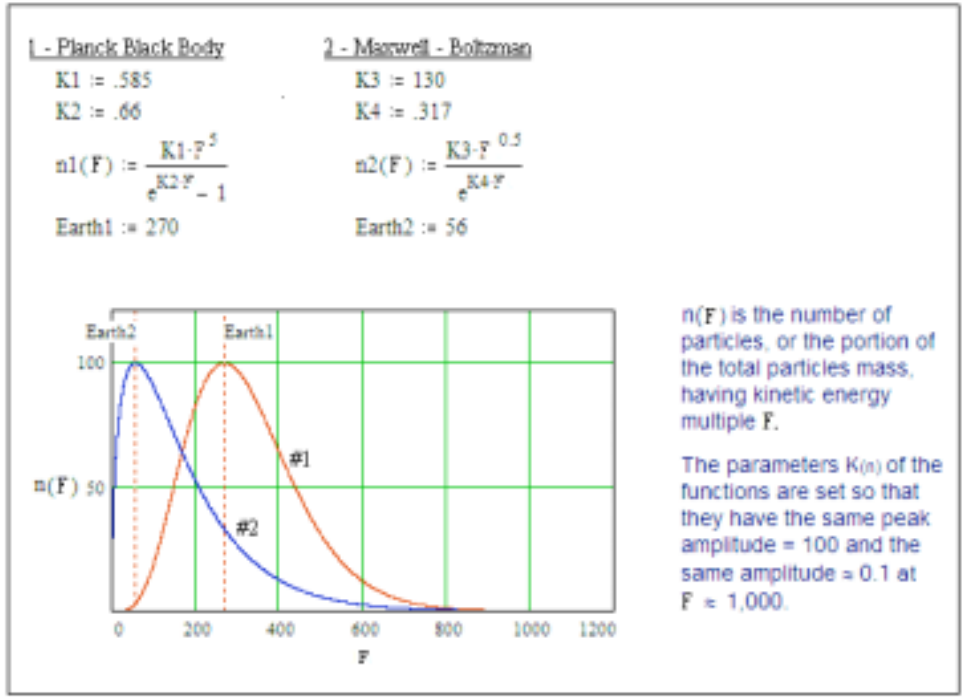

*Figure 1a -- "Big Bang": Some Theoretical Rate Distributions of Initial Particle Energies*

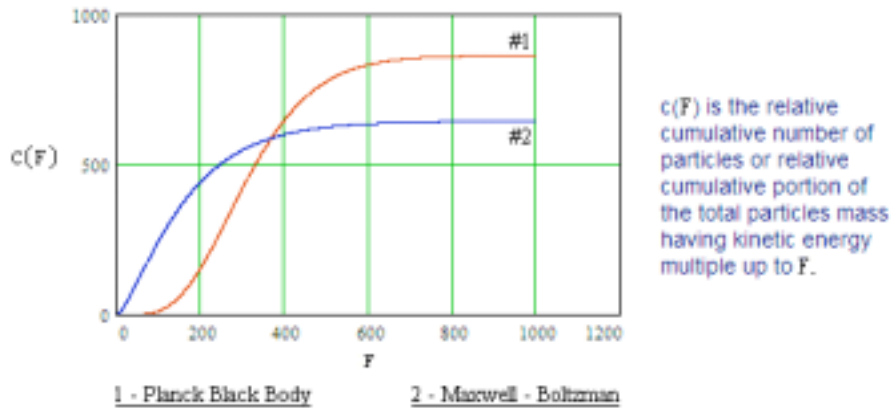

*Figure 1b -- Cumulative Distributions for Figure 1a*

Where we observers on planet Earth fall in the distributions in the above Figure 1a should be considered. It can only be presumed that Earth is not unusual with regard to its component particles' initial velocities, which would call for placing it at a distribution peak. But, there are two choices shown, likely neither is exact, and Earth is not necessarily so usual as to fall exactly at the peak of any distribution. Form #2 requires less total energy and also is chosen because otherwise the resulting velocities at `Age = 30 Gyrs` would appear to be in error relative to the known velocity of Earth (see before equation *(27)*, further below. Calculating for the two cases in Figure 1a shows that the variation in choices produces little variation in the overall results and in the age, `30 Gyrs`, of the universe and the theoretical limit, about `27 Gyrs`, on how far back in the past can be observed. The Earth case parameter, `F = 55`, [see Tables 2a and 2b, further below] is chosen to set it at the distribution Form #2 peak.

From the above data the summary conclusions in Table 1c, below, can be drawn.

| Form #1 -- Planck Black Body Radiation | | | |
|---|---|---|---|
| Percent of Maximum Amplitude, `n(F)`: | `100 %` | `95 %` | `90 %` |
| Range in Distribution is `F = 0 to`: | `1,000` | `565` | `500` |
| That Range as Percent of the Maximum | `100 %` | `57 %` | `50 %` |
| *Resulting Indicated Universe Radius* | `~14 G-Lt-Yrs` | `~14 G-Lt-Yrs` | `~14 G-Lt-Yrs` |
| Form #2 -- Maxwell - Boltzman Gas Kinetic Theory | | | |
| Percent of Maximum Amplitude, `n(F)`: | `100 %` | `95 %` | `90 %` |
| Range in Distribution is `F = 0 to`: | `1,000` | `455` | `360` |
| That Range as Percent of the Maximum | `100 %` | `46 %` | `36 %` |
| *Resulting Indicated Universe Radius* | `~14 G-Lt-Yrs` | `~14 G-Lt-Yrs` | `~14 G-Lt-Yrs` |

*Table 1c -- "Big Bang" Initial Energy Distributions Summary Data Conclusions*

In the above table, the "*Resulting Indicated Universe Radius*" in `G-Lt-Yrs` is obtained as follows. Figure 4d, further on below, *Second Phase of The Expansion of The Universe -- Distances for Age = 30 Gyrs Case* indicates a present [at `Age of the universe ≈ 30 Gyrs`] radius of the matter-containing volume of the universe as about `8 G-Lt-Yrs`. However, the radius applicable to the above obtained universe mass density should be based on an earlier time because the investigations into estimating that density had to treat astral objects which we observe as they were some time in the past: their distance from us divided by the speed of their light. Taking those earlier times as having been in the range of `0 to 7-8 Gyrs` into the past, which corresponds to volumes in the ratio to each other of the cube of those distances as `[0, 1, 8, 27, 64, 125, 216, 343, 512]`, and cumulatively in ratio as `[0, 1, 9, 36, 100, 225, 441, 784, 1296]` then it is reasonable to take the applicable universe radius as that which existed at the time into the past corresponding to about half the maximum cumulative volume, `t ≈ 6.5 Gyrs` ago. Figure 4d indicates the radii given in the above Table 1c for the related table columns at that time ago, `≈ 14 G-Lt-Yrs`.

Then, the estimated radius of the universe for the present calculation is:

*(15)*    `R_U = 14 G-Lt-Yrs`

        `= 11·10^24 meters.`

Therefore the mass of the universe, as the product of its volume based on that radius and its equation *(13)* density, is:

(16)   $m_U = 3 \cdot 10^{49}$ kg.

[Calculating with alternative values for the mass of the universe ranging from $10^{46}$ to $10^{53}$ kg produces no significant change in the general results developed below as can be verified using the forms of the calculations presented further on below. That is, while velocities and distances vary somewhat, the necessary age of the universe remains at about the 30 Gyrs and the maximum distance back into the past that it is theoretically possible to observe remains at about the 27 Gyrs developed further on below.]

[A possible concern over circular cause and effect reasoning here is not valid. The results presented are based on numerous iterations of calculations over a range of complexly interacting variables.]

With the mass of the universe now resolved the other quantities needed to calculate the escape velocity of the universe can be addressed. The <u>Separation Distance</u>, $d_0$, is the radius of the universe at the moment that expansion began being at a rate consistent with the long term development of the universe as compared to its initial more rapid [essentially instantaneous] development commonly referred to as "inflation". That value is $d_0 = 4.0 \cdot 10^7$ meters.

$G(0)$, the <u>Gravitation Constant</u> at its original un-decayed value at the time of the "Big Bang" is as follows.

(17)   $G(t) = G(0) \cdot \varepsilon^{-3 \cdot t/\tau}$   $m^3/kg \cdot s^2$    [Form per the Universal Decay and the [$m^3$] requires $\tau \to \tau/3$]

$G(Age) = 6.672,59 \cdot 10^{-11}$   $m^3/kg \cdot s^2$    ["Our" G, now]

$G(Age) = G(0) \cdot \varepsilon^{-3 \cdot Age/\tau}$    [Set t = Age in G(t)]

$G(0) = G(Age) \cdot \varepsilon^{+3 \cdot Age/\tau}$    [Solve for G(0)]

      $= 6.672,59 \cdot 10^{-11} \cdot \varepsilon^{3 \cdot Age/\tau}$   $m^3/kg \cdot s^2$

Then, the gravitation constant at any arbitrary time, $t$, after the "Big Bang" for any arbitrary age of the universe, $Age$, is as follows.

(18)   $G(t, Age) = G(0) \cdot \varepsilon^{-3 \cdot t/\tau}$

          $= [6.672,59 \cdot 10^{-11} \cdot \varepsilon^{3 \cdot Age/\tau}] \cdot \varepsilon^{-3 \cdot t/\tau}$   meters/sec

Two values for the $Age$ of the universe are addressed in this analysis to present the thesis and its validation. The currently accepted values in Hubble - Einstein cosmology range $Age = 13.5$ to $14.7$ Gyrs. Representing those $14.0$ Gyrs will be used. As developed below, the present analysis indicates that $Age = 30.0$ Gyrs. Then, using $\tau = 11.3373$ Gyrs from Equation (2) the following values for $G(0)$ result.

(19)   <u>For Age = 14 Gyrs</u>      <u>For Age = 30 Gyrs</u>

     $G(0) = 2.711 \cdot 10^{-9}$      $G(0) = 1.870 \cdot 10^{-7}$

The escape velocity per equation (8) for those cases of age of the universe are:

(20)   <u>For Age 14 Gyrs</u>      <u>For Age 30 Gyrs</u>

     $v_{esc} = 6.4 \cdot 10^{16}$ m/s      $v_{esc} = 5.3 \cdot 10^{17}$ m/s

Those values are so large relative to the speed of light at the time of the "Big Bang",

(21)   <u>For Age 14 Gyrs</u>      <u>For Age 30 Gyrs</u>

     $c(0) = 1.031 \cdot 10^9$ m/s      $c(0) = 4.227 \cdot 10^9$ m/s

that it is certain that the initial particle velocities, at the time of the "Big Bang", were very nearly the then speed of light. That is, the initial particle velocities could not be, nor exceed, light speed as the non-relativistically calculated escape velocities of equation *(20)* call for. The accommodation to relativity means that the actual speeds were very near light speed and the masses were significantly relativistically increased.

As noted earlier, to determine the travel of cosmic bodies outward from the "Big Bang" one needs to know, first, the initial velocities and then the subsequent manner of the reduction in the cosmic bodies' velocities by gravitation and other effects. The initial velocities have been found to be essentially the value of the speed of light at the time of the "Big Bang". At that point two effects proceeded to slow the outward velocities: the decay of the speed of light [the upper limit on particle velocity] and the gravitational slowing [the centrally directed gravitational acceleration caused by the total mass].

### *(2) The Progressive Reduction in the Cosmic Bodies' Initial Velocities*

The overall process must be divided into two phases:

- First, the relativistic phase during which the speed is continuously almost that of light and the effect of gravitation is dominantly not a reducing of the speed but a reducing of the amount that the mass has been relativistically increased, and

- Second, the non-relativistic phase during which the mass, now reduced to essentially rest mass, remains essentially the same and the dominant effect of gravitation is to reduce the speed.

Of course the change from the first to the second phase is not sharp, but rather a gradual smooth transition. For the purposes of these calculations, however, the choosing of a specific transition point [hereafter termed the `ChangePoint`] is needed. That point is determined as follows.

### *(2a) The First Calculation Phase -- Speed ≈ Light Speed*

The first phase calculation is in terms of energy, the gradual transfer of initial kinetic energy into gravitational potential energy. Energy calculations in themselves are not relativistic. The kinetic energy speed will be treated non-relativistically, that is, the mass is taken as at its rest value and the kinetic energy then is taken as all residing in the [theoretical] velocity squared, that theoretical velocity not constrained by a speed of light limitation. Then, the calculated effect of gravitation, of the transfer of kinetic energy into gravitational potential energy, appears as a gradual reduction of that theoretical velocity. When that theoretical velocity has been reduced by gravitation down to the actual [at that time as decayed] light speed then the `ChangePoint` from the relativistic to the non-relativistic treatment has been reached.

During that first phase the distance component of the gravitational potential energy calculation is readily available as the time integral of the known speed, the speed of light. The velocity as a function of time then develops as follows.

The first phase distance, $d(t,Age)$, traveled outward from the "Big Bang" source location, as a function of time is the time integral of the velocity as equation *(21)*, below, which is based on equation *(7)*, above [and includes the initial separation distance, $d_0 = 4.0 \cdot 10^7$ *meters*, of the earlier above calculation of the mass of the universe, which distance is negligible, however].

*(21)*
$$d(t,Age) = d_0 + \int_0^t c(t) \cdot dt$$

$$= 4.0 \cdot 10^7 + \int_0^t \left[ 2.997{,}924{,}58 \cdot 10^8 \cdot \varepsilon^{Age/\tau} \right] \cdot \varepsilon^{-t/\tau} \cdot dt$$

The velocity as a function of time, $v(t,Age)$, is, starting from equation *(8)*:

(22)
$$v_{esc} = \left[\frac{2G \cdot m_U}{d_0}\right]^{1/2} \quad \text{so that:}$$

$$v(t,Age) = \left[\frac{2G \cdot m_U}{d(t,Age)}\right]^{1/2}$$

As pointed out during the evaluation of the mass of the universe earlier above, the matter of the universe moved outward from the "Big Bang" at a wide range of speeds. Those various speeds resulted, of course, from the particles of matter having various initial velocities / energies which, as presented just before and in conjunction with Figures 1a and 1b, are to be sampled over the range $F = 1$ to $1,000 \times$ *[the escape energy]*. That range is incorporated into the formulation by the multiple factor, $F$, included in the final expression for the first phase $v1(t,Age)$ per below.

(23)
$$v1(t,Age) = \left[\frac{\mathbf{F} \cdot 2G \cdot m_U}{d(t,Age)}\right]^{1/2} \quad \text{[1st Phase Matter Velocities]}$$

The decaying speed of light is per equation *(7)*, repeated below,

(24)
$$c(t,Age) = [2.997,924,58 \cdot 10^8 \cdot \varepsilon^{Age/\tau}] \cdot \varepsilon^{-t/\tau}$$

and the value of time, $t$, producing $\boxed{v1(t,Age) \equiv c(t,Age)}$ is the sought *ChangePoint* for the particular initial velocity / energy multiple factor, $F$, and $Age$, the end of calculation for the first, the relativistic, phase.

Tables 2a and 2b, below summarize the results for the first phase for both *Age = 30* and *Age = 14 Gyrs*, and the results are also presented graphically for *Age = 30 Gyrs* in Figure 2c, following the tables.

```
For:   Universe Age = 30 Gyrs, which means that:
       Initial Light Speed = 4.226,895,62·10⁹ m/s
       Initial Gravitation Constant, G = 1.870,24·10⁻⁷ m³/kg-s²
```

| | | "ChangePoint" | | | |
|---|---|---|---|---|---|
| F-<br>Factor | At<br>Time[Gyrs] | At<br>Velocity[m/s] | Distance From Origin*<br>ChangePoint    Now, Age | | Relative!<br>Abundance |
| 1 | 0.004713 | 4.225·10⁹ | 0.066 | 0.005 | 22. |
| 3 | 0.01414 | 4.222·10⁹ | 0.199 | 0.014 | 37. |
| 10 | 0.04721 | 4.209·10⁹ | 0.661 | 0.047 | 63. |
| 32 | 0.1518 | 4.171·10⁹ | 2.097 | 0.152 | 93. |
| 55 Earth | 0.2621 | 4.131·10⁹ | 3.568 | 0.262 | 100. |
| 100 | 0.4812 | 4.052·10⁹ | 6.364 | 0.481 | 90. |
| 316 | 1.5960 | 3.672·10⁹ | 18.222 | 1.596 | 23. |
| 1000 | 6.0840 | 2.472·10⁹ | 38.788 | 6.084 | 0.1 |
| ≈3000 | → ∞ | c(t) | → ∞ | → ∞ | |

```
* = Decayed to Change Point, Age; G-Lt-Yrs.    ! = Estimate per Figure 1a
```

*Table 2a - The Universe's First Phase of Expansion, Age=30 Gyrs Case --*
*The Relativistic Phase at (Essentially) Light Speed, From t = 0 to t = ChangePoint*

```
For:  Universe Age = 14 Gyrs, which means that:
      Initial Light Speed = 1.030,357,62·10^9 m/s
      Initial Gravitation Constant, G = 2.711,29·10^-9 m^3/kg-s^2
```

| F-Factor | At Time[Gyrs] | At Velocity[m/s] | "ChangePoint" Distance From Origin* ChangePoint | Now, Age | Relative! Abundance |
|---|---|---|---|---|---|
| 1 | 0.004715 | 1.030·10^9 | 0.016 | 0.005 | 22. |
| 3 | 0.01416 | 1.029·10^9 | 0.049 | 0.014 | 37. |
| 10 | 0.04721 | 1.026·10^9 | 0.161 | 0.047 | 63. |
| 32 | 0.1518 | 1.017·10^9 | 0.511 | 0.151 | 93. |
| 55 Earth | 0.2621 | 1.007·10^9 | 0.870 | 0.259 | 100. |
| 100 | 0.48125 | 9.878·10^8 | 1.551 | 0.471 | 90. |
| 316 | 1.5961 | 8.953·10^8 | 4.443 | 1.488 | 23. |
| 1000 | 6.089 | 6.024·10^8 | 9.460 | 4.708 | 0.1 |
| ≈3000 | → ∞ | c(t) | → ∞ | → ∞ | |

\* = Decayed to Change Point, Age; G-Lt-Yrs.    ! = Estimate per Figure 1a

*Table 2b - The Universe's First Phase of Expansion, Age=14 Gyrs Case --*
*The Relativistic Phase at (Essentially) Light Speed, From t = 0 to t = ChangePoint*

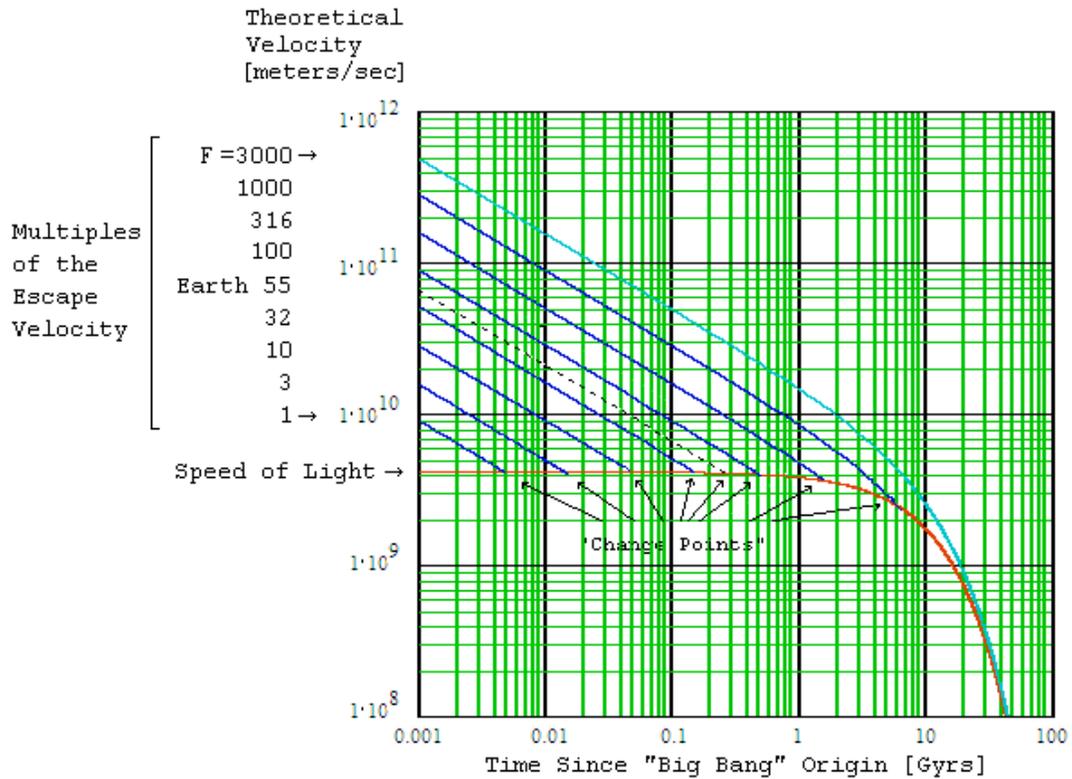

*Figure 2c*
*First Phase of The Expansion of The Universe -- Velocities for Age = 30 Gyrs Case*

Note that for values of the `F-Factor` at about `F = 3,000` and above the "*ChangePoint*" is never reached because of the decay in the speed of light. For those values the outward moving matter never slows below, essentially, the then on-going decaying light speed.

Note, also, that the large Doppler red shift resulting from *v* nearly equaling *c* combined with large decay redshift due to lack of much decay because the time lies only shortly after *t = 0* results in a redshift relative to our "normal" local wavelengths by a factor of *24 – 28*. The least wavelength of visible light is about *0.38 microns*. That shifted to *0.38 x [24 – 28] = [9 – 11] microns*, lies well into the infra-red portion of the spectrum.

Consequently, the light emitted from sources before they reach their "ChangePoints", which light would otherwise lie in the "visible light" portion of the spectrum, lies shifted sufficiently into the infra-red that its detection is relatively unlikely, especially since sources with relatively later "ChangePoints" [more recent, therefore more susceptible to observation] are of relatively small relative abundance. In other words, light emitted from astral sources before they reached their *ChangePoint* is much less likely to be observed.

### (2b) *The Second Calculation Phase -- Speed < Light Speed*

Gravitational slowing is an awkward problem. The amount of gravitational slowing depends on the distance outward from the origin of the "Big Bang"; those distances depend on the velocity function during the travel from the origin outward; and that velocity function depends on the gravitational slowing -- a problem of circular cause and effect. The calculation breaks down into two different modes of behavior because of relativistic effects.

The first phase of the outward expansion, already analyzed above, takes place at essentially the actual speed of light regardless of gravitation. That is because the [theoretical non-relativistic] escape velocities are so large. The kinetic energy essentially resides in the relativistically increased mass until gravitation has reduced the [theoretical non-relativistic] greater than light speed down to passing through and below the actual light speed. During that first phase distances are known because the velocities are known independently of the distance; they are essentially the then current, as decayed, light speeds.

The second phase begins at the end of that first phase's known outward travel to its "*ChangePoint*". At that point the circular cause and effect awkwardness of the problem returns. The inverse square gravitational behavior calls for the current total outward distance squared in its denominator and that depends on the velocity history which depends on the distance history which depends on the velocity history. The solution is to use an approximating function of similar form but not involving velocity. That function can then be adjusted by calibrating it to the known velocity of the Earth now, as developed further below.

The <u>approximating function</u> develops as follows. The precise behavior of the universe's matter expanding outward from the "Big Bang" is as equation *(25)*, the distance represented by the variable *s* to avoid confusion with the symbols for differentiation.

*(25)*
$$\text{Gravitational Deceleration} = \frac{d^2 s}{dt^2} = - \frac{G \cdot \text{UniverseMass}}{s^2}$$

The general form of the solution to that equation is as equation *(26)*

*(26)*
$$\text{Velocity} = \frac{ds}{dt} = \frac{1}{A \cdot \varepsilon^{B \cdot s} + C}$$

which states that the velocity is inversely proportional to the exponential of the distance, *s*, as $1/\varepsilon^s$. The procedure will be to use as the approximation to the actual exact velocity function the function: the speed of light, c(t), per equation *(24)*, multiplied by a factor based on $1/\varepsilon^s$.

However, the velocity function must be in terms of time, *t*, as the independent variable, not distance, *s*, else the problem of circular cause and effect remains. Using *t* instead of *s*, that is representing the actual exact velocity function with a function multiplying the speed of light, *c(t)*, by a factor based on $1/\varepsilon^t$ resolves that problem but is less accurate an approximation. The problem of accuracy is addressed by calibrating to the known velocity of our planet Earth at time the  Age  of the universe.

Doppler analysis of the cosmic microwave background radiation shows that the absolute velocity of the Earth [absolute relative to the location of the "Big Bang" origin] is now about $3.7 \cdot 10^5$ meters/sec. The calibration of the velocity approximating function must be such as to produce approximately that velocity now, at time $t$ = Age after the "Big Bang"; but, for what case, what value of the  F-Factor ? That issue has been already addressed above and the result is that, for *Earth* F = 55 will be used.

So that the beginning of the second phase will match smoothly with the ending of the first phase the adjustment, $1/_\varepsilon t$, which increases in its effect as $t$ increases, must be formulated to produce zero change when the time is $t$ = ChangePoint. The resulting expression for the second phase velocity is equation (27), below.

(27)
$$v2(t, Age) = c(t, Age) \cdot \frac{1}{\varepsilon^{A \cdot (t - ChangePoint)}} \quad \text{meters/sec}$$

where A is a constant of value yet to be determined, chosen to
calibrate the function, and t must always be t ≥ ChangePoint.

The calibrating constant, A, for the case of *Earth*, F-Factor = 55 is set to the value that produces a velocity at Age of $3.7 \cdot 10^5$ meters/sec, the known value. The A for the highest  F-Factor  case is set to produce a velocity of the current speed of light, $3 \cdot 10^8$ meters/sec, at Age. The A for each of the other cases is interpolated using a decaying exponential form to provide a general representative set of samples. That form is per equation (28) and as depicted in Figure 3 below it.

(28)  $Velocity(i) = 5.452 \cdot 10^9 \cdot [\varepsilon^{-2.2127(9-i)}]$

where i = case # = 1, 2, ... 8

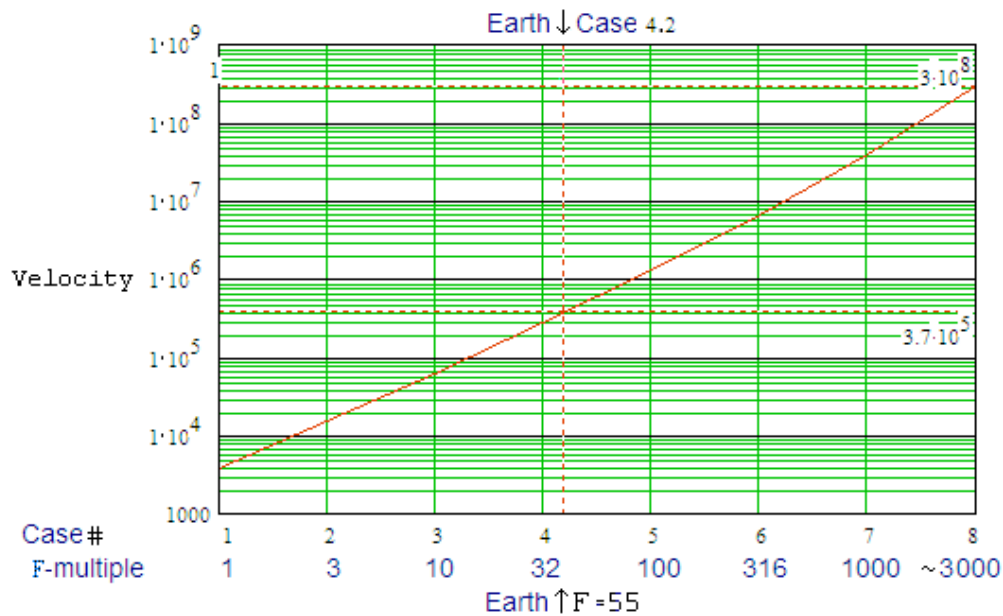

*Figure 3*
*Calibrated Velocities at "Age" for Sample Expansion Cases #1 - #7, and #Earth*

The development of the formulations for the distances is presented further below at equations (31A) through (33B). Tables 4a and 4b below summarize the results for the second phase for *Age = 30 Gyrs* and for *Age = 14 Gyrs*, and the results are also presented graphically for *Age = 30 Gyrs* in Figure 4c, *Second Phase of The Expansion of The Universe -- Velocities for Age = 30 Gyrs Case*, and Figure 4d, *Second Phase of The Expansion of The Universe -- Distances for Age = 30 Gyrs Case*, on the page following the tables.

```
    For:  Universe Age = 30 Gyrs, which means that:
          Initial Light Speed = 4.226,895,62·10⁹ m/s
          Initial Gravitation Constant, G = 1.870,24·10⁻⁷ m³/kg-s²
```

| F-Factor | ChangePoint Time[Gyrs] | 2nd Phase Constant-A | At Age = 30 Gyrs,[Data*] Velocity [m/s] | Distance[G-Lt-Yrs] |
|---|---|---|---|---|
| 1 | 0.004713 | 16.92385 | 0.00003814·10⁸ | 2.157 |
| 3 | 0.01414 | 16.98010 | 0.0001505 ·10⁸ | 2.400 |
| 10 | 0.04721 | 17.04691 | 0.0006230 ·10⁸ | 2.727 |
| 32 | 0.1518 | 17.12856 | 0.002731 ·10⁸ | 3.205 |
| 55 | 0.2621 | 17.14622 | 0.003700 ·10⁸ | 3.374 |
| 100 | 0.4812 | 17.23245 | 0.01287 ·10⁸ | 3.980 |
| 316 | 1.5960 | 17.37183 | 0.06673 ·10⁸ | 5.388 |
| 1000 | 6.0840 | 17.57200 | 0.3974 ·10⁸ | 8.034 |
| ≈3000 | → ∞ | n/a | 2.99792458·10⁸ | 10.526 |

\* = Decayed to Age

*Table 4a - For Age = 30 Gyrs*
*Summary Data Results for the Universe's Second Phase of Expansion --*
*The Non-Relativistic Phase From t = "ChangePoint" Onward*

```
For:  Universe Age = 14 Gyrs, which means that:
      Initial Light Speed = 1.030,357,62·10⁹ m/s
      Initial Gravitation Constant, G = 2.711,29·10⁻⁹ m³/kg-s²
```

| F-Factor | ChangePoint Time[Gyrs] | 2nd Phase Constant-A | At Age = 14 Gyrs,[Data*] Velocity [m/s] | Distance[G-Lt-Yrs] |
|---|---|---|---|---|
| 1 | 0.004715 | 16.59278 | 0.00003814·10⁸ | 1.122 |
| 3 | 0.01416 | 16.64887 | 0.0001505 ·10⁸ | 1.268 |
| 10 | 0.04721 | 16.71513 | 0.000623 ·10⁸ | 1.477 |
| 32 | 0.1518 | 16.79505 | 0.002731 ·10⁸ | 1.811 |
| 55 | 0.2621 | 16.81082 | 0.003700 ·10⁸ | 1.954 |
| 100 | 0.4812 | 16.89330 | 0.01287 ·10⁸ | 2.417 |
| 316 | 1.5961 | 17.01200 | 0.06673 ·10⁸ | 3.669 |
| 1000 | 6.0890 | 17.09155 | 0.3974 ·10⁸ | 6.295 |
| ≈3000 | → ∞ | n/a | 2.99792458·10⁸ | 8.034 |

\* = Decayed to Age

*Table 4b - For Age = 14 Gyrs*
*Summary Data Results for the Universe's Second Phase of Expansion --*
*The Non-Relativistic Phase From t = "ChangePoint" Onward*

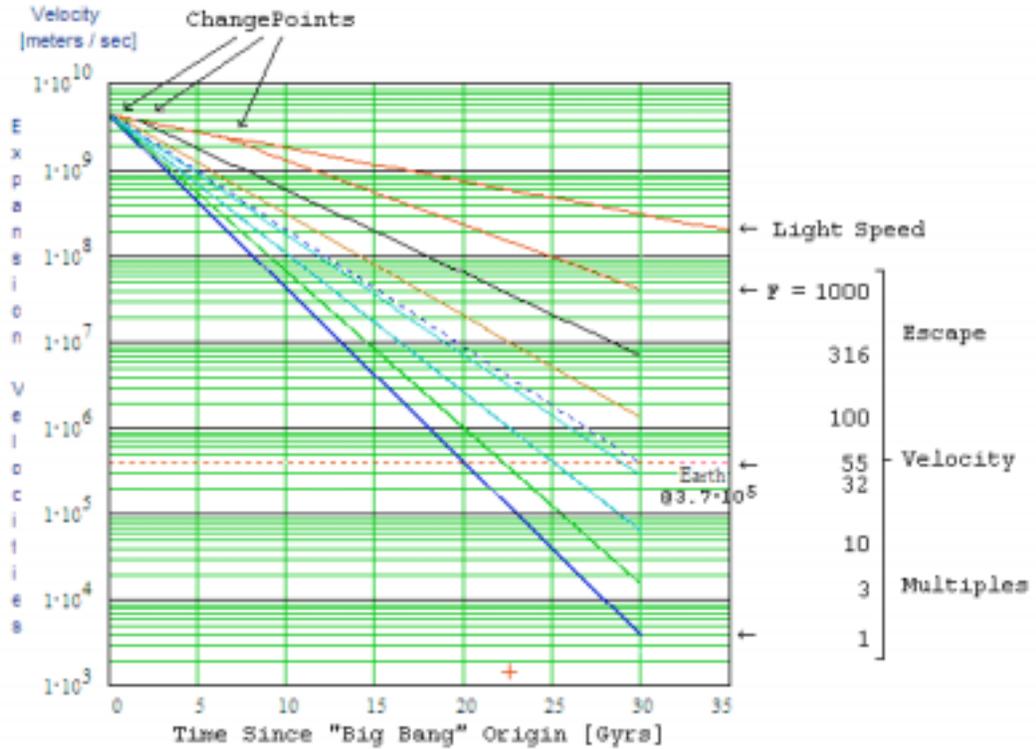

*Figure 4c*
*Second Phase of The Expansion of The Universe -- Velocities for Age = 30 Gyrs Case*

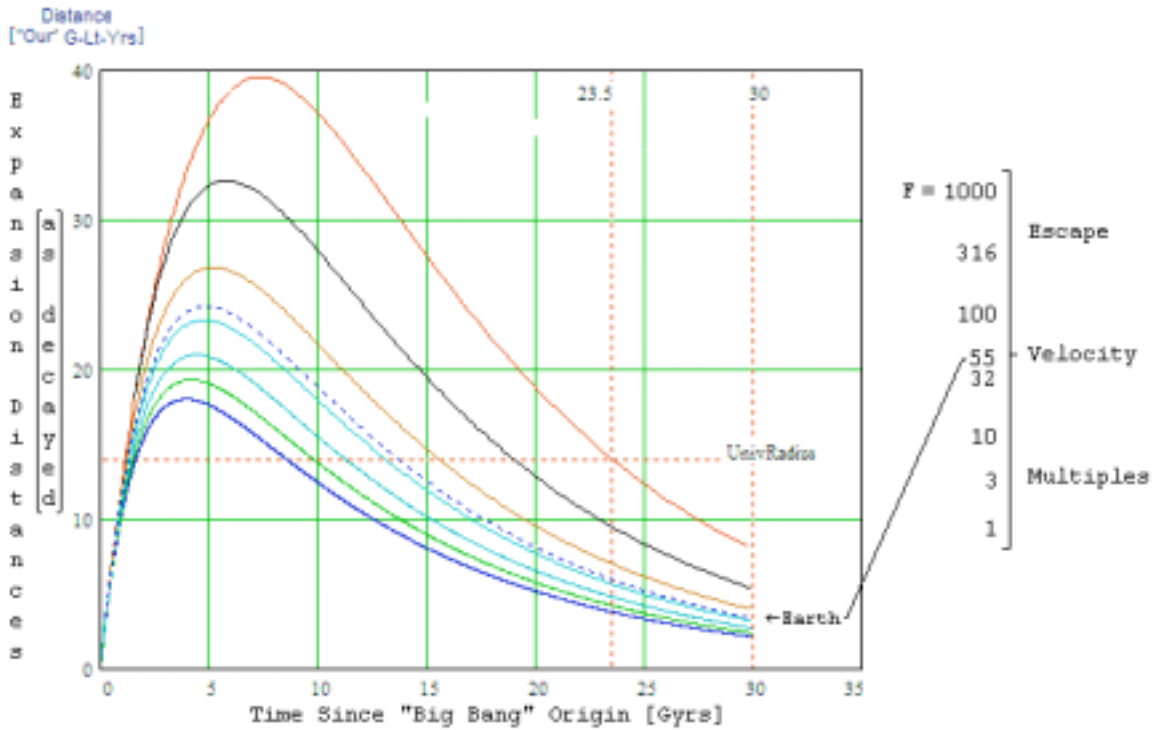

*Figure 4d*
*Second Phase of The Expansion of The Universe -- Distances for Age = 30 Gyrs Case*

[The notation "UnivRadius" in the above Figure 4d refers to the discussion following Table 1c concerning developing the appropriate volume of the universe to use in conjunction with the universe's density to obtain the universe's mass.]

*2 - Observing, from Our Planet Earth, Light Emitted by Astral Sources*

Now that formulations have been developed that describe the travel of the two different traveling entities, light and matter, at various times in the past from at the beginning to the present the problem of observing light from astral sources can be addressed. The problem is to determine under what conditions light emitted long ago will have traveled to the exact present location of an observer that has also been performing its own travel while the light to be observed has been traveling.

For convenience the following quantities are defined.

```
Age  ≡ Age now of the universe = time "now".
Back ≡ how long ago it is theoretically possible to observe.
Then ≡ the Age of the universe then [Back ago].
     = Age - Back.
```

The distance that the light travels [continuously at whatever its speed was when it was first emitted from its astral source] from when first emitted at time $t$ until now, at time *Age*, is then:

*(29)*   `LightTravels(t,Age)` = [Speed(t)]·[Travel Time]

                       = [c(t,Age)[m/s]]·[[Age - t][s]]

                      = c(t,Age)·[Age - t]     meters

Consistent with the mnemonic terminology above, `LightTravels(t,Age)`, the motion of the cosmic bodies involved will be termed as in equation *(30)*.

*(30)*   `WhereUs(t,Age,F)`    ≡ location of we observers

        `WhereSource(t,Age,F)` ≡ location of source of the observed light

                             ≈ -WhereUs(t,Age,F)

```
to account for the formulations being of similar form except that
that their travels are diametrically opposite for the purposes of
the present calculations and their "F" values can be different.
```

Taking the location of the "Big Bang" singularity as at distance zero and noting that the initial distance, $d_0 = 4.0 \cdot 10^7$ *meters*, is negligible [less than one "our light years" on a scale of "giga our light years"], then the formulation for `WhereUs(t,Age)` develops as follows.

*(31)*   `Travel from the "Big Bang" outward`

 *(31A)* `Travel during time t=0 to time t=ChangePoint:`

       `WhereUsA(t,Age,F)` = [Travel of Matter in 1st Phase]

$$= \int_0^t [\text{Speed of Matter in 1st Phase}(t,Age,F)] \cdot dt$$

$$= \int_0^t [\text{Decaying Light Speed}(t,Age)] \cdot dt$$

$$= \int_0^t c(t,Age) \cdot dt \quad \text{meters}$$

*(31B)* Travel from time t=ChangePoint onward:

```
WhereUsB(t,Age,F) =
```

$$= [\text{1st Phase Travel to ChangePoint}] + [\text{2nd Phase Travel}(t,Age,F)]$$

$$= [\text{WhereUsA(ChangePoint,Age)}] + [\text{2nd Phase Travel}(t,Age,A)] \qquad [F \to A]$$

$$= [\text{1st Term}] + \int_{\text{ChangePoint}}^{t} [\text{DecayingLightSpeed}] \cdot [\text{GravitySlowing}] \cdot dt$$

$$= [\text{1st Term}] + \int_{\text{ChangePoint}}^{t} [c(t)] \cdot \frac{1}{\varepsilon^{A \cdot [t - \text{ChangePoint}]}} \cdot dt \qquad \text{meters}$$

and where "A" is the "2nd Phase Constant-A" of Tables 4a and 4b.

However, there is one more factor in the development of *WhereUs(t,Age)*, the general effect of the Universal Decay. The Universal Decay produces an acceleration on all bodies which acceleration is: centrally directed, independent of separation distances, and the same and constant everywhere except that the amount of that acceleration also exponentially decays. Its value now is *(8.74 ± 1.33) × 10$^{-8}$ cm/s2 $^2$*. That acceleration produces a gradual contraction of the overall universe; that is, the exponential decay of the length *[L]* aspect of all quantities is also a decay of the distance spacings in the universe.

The acceleration is evidenced by galactic rotation curves and by the travel of the Pioneer 10 and 11 space craft. That the Universal Decay is not directly observable because our measuring equipment [our "ruler"] is also decaying, as noted earlier above, prevents our direct observation of the contraction in the case of our solar system and that of galactic rotation curves. However, in the case of the Pioneer 10 and 11 satellites and the case of galactic rotation curves, the decay has been detected because it forces orbital / path behavior that would not be present if there were no decay, and that behavior can be and has been observed -- e.g. the Pioneer space craft are not as far outward from the Sun as they should be were there no Universal Decay contraction.

Consequently, as the various bodies in the universe travel outward from the location of the "Big Bang", the distances that they have already traveled continuously decay. And, consequently, the distances traveled by light emitted from the various sources in the universe continuously decay after having been first traveled "undecayed". Therefore, the final form for *LightTravels* is then equation *(32)*, below [continued from equation *(29)*]. And, the final form for *WhereUs(t,Age)* is equations *(33A)* and *(33B)*, below [continued from equations *(31A)* and *(31B)*, above].

*(32)*     Distance traveled outward from its source until
          now, time "Age", by light emitted at time "t":

$$\text{LightTravels}(t,\text{Age}) = [\text{Speed}] \cdot [\text{Travel Time}] \cdot [\textit{As Decayed}]$$

$$= c(t,\text{Age}) \cdot [\text{Age} - t] \cdot \varepsilon^{-[(\text{Age}-t)/\tau]} \qquad \text{meters}$$

*(33)*     Distance traveled outward from the "Big Bang" until time "t"
          by matter originating at the "Big Bang" [t=0, distance=0]:

WhereUs(t,Age,F):

(33A): Travel during time t=0 to time t=ChangePoint:

$$\text{WhereUsA}(t,\text{Age},F) = [\text{Travel of Matter in 1st Phase}] \cdot [\textit{As Decayed}]$$

$$= \left[ \int_0^t c(t,\text{Age}) \cdot dt \right] \cdot \varepsilon^{-[t/\tau]} \qquad \text{meters}$$

*(33B):* Travel from time t=ChangePoint onward:

WhereUsB(t,Age,F) =

= [[1st Phase Travel to ChangePoint]·[not decayed] +
         + [2nd Phase Travel(t,Age,A)]]·*[All as Decayed]*

$$= \left[ [\text{1st Term}] + \int_{\text{ChangePoint}}^{t} [c(t)] \cdot \frac{1}{\varepsilon^{A \cdot [t - \text{ChangePoint}]}} \cdot dt \right] \cdot \varepsilon^{-[t/\tau]} \text{ meters}$$

### a. *Calculation of the Maximum Distance into the Past That is Observable*

The extreme case of observing ancient light is the observing of light that originated diametrically opposite from us, the observers, relative to the origin of the "Big Bang". That light <u>must</u> travel the distance from its source back to the location of the origin of the "Big Bang" and then further outward to the location of us, the observers. The light originates from its source at time $t = \text{Then}$ and is observed by us at time $t = \text{Age}$. That distance, for any age of the universe, $\text{Age}$, is as follows.

*(34)*  LightMustTravel(t,Age)
                     = -WhereSource(Then,Age,F) + WhereUs(Age,Age,F)

As compared to the above requirement, the actual distance that that light <u>does</u> travel is given by equation *(32)* with $t = \text{Then}$, as follows.

*(35)*  LightDoesTravel(t,Age) = c(Then,Age)·[Age - Then]·$\varepsilon^{-[(\text{Age} - \text{Then})/\tau]}$

           = c(Then,Age)·[Back]·$\varepsilon^{-[(\text{Back})/\tau]}$   meters

For the light to be theoretically observable by us the above two must be the same.

*(36)*  LightMustTravel(t,Age,F) = LightDoesTravel(t,Age)

The only variables in equation *(36)* [for a particular $\text{Age}$ and energy multiple, $F$,] are $\text{Back}$ and $\text{Then}$, either of which determines the other per $\text{Then} = \text{Age} - \text{Back}$. The solution to equation *(36)* is obtained using a computer assisted design program ["Mathcad" in this case]. The applicable form of WhereUs(t,Age,F) must be used, WhereUs<u>A</u>(t,Age,F) or WhereUs<u>B</u>(t,Age,F) depending on the value of $\text{Then}$ relative to the $\text{ChangePoint}$.

The results are presented in Tables 5a and 5b, below.

```
For:  Universe Age = 30 Gyrs, which means that:
      Initial Light Speed = 4,226,895.62·10^9 m/s
      Initial Gravitation Constant, G = 1.870,24·10^-7 m^3/kg-s^2
```

| F-Factor | ChangePoint Time[Gyrs] | 2nd Phase Constant-A | Observable Past Distance Back [Gyrs] | Observable Past Distance Then [Gyrs] | Relative! Abundance |
|---|---|---|---|---|---|
| 1 | 0.004713 | 16.92385 | 7.5860 | 22.4140 | 22. |
| 3 | 0.01414 | 16.98010 | 8.4081 | 21.5919 | 37. |
| 10 | 0.04721 | 17.04691 | 9.8790 | 20.1210 | 63. |
| 32 | 0.1518 | 17.12856 | 26.0791 | 3.9209 | 93. |
| 100 | 0.4812 | 17.23245 | 27.01869 | 2.98131 | 90. |
| 316 | 1.5960 | 17.37183 | 27.56523 | 2.43477 | 23. |
| 1000 | 6.0840 | 17.57200 | 27.61821 | 2.38179 | 0.1 |

! = Estimate per Figure 1a, where Earth, F = 55, is of Abundance 100

*Table 5a*
*Distance into the Past That is Observable, Age = 30 Gyrs*

```
For:  Universe Age = 14 Gyrs, which means that:
      Initial Light Speed = 1.030,357,62·10^9 m/s
       Initial Gravitation Constant, G = 2.711,29·10^-9 m^3/kg-s^2
     F-      ChangePoint    2nd Phase    Observable Past Distance    Relative!
   Factor   Time[Gyrs]    Constant-A    Back [Gyrs]    Then [Gyrs]   Abundance
     1        0.004715      16.59278       3.4828        10.5172        22.
     3        0.01416       16.64877       3.7136        10.2864        37.
    10        0.04721       16.71513       4.0702         9.9298        63.
    32        0.1518        16.79505       4.6849         9.3151        93.
   100        0.48125       16.89330       5.9772         8.0228        90.
   316        1.5961        17.01200       8.8230         5.1770        23.
  1000        6.0890        17.09155      10.04945        3.95055        0.1
       ! = Estimate per Figure 1a, where Earth, F = 55, is of Abundance 100
```

*Table 5b*
*Distance into the Past That is Observable, Age = 14 Gyrs*

For *Age = 14 Gyrs*, as presented in Table 5b above, even the most energetic case of observable past distance, that for *F = 1,000*, has a theoretical limit, about *10 Gyrs*, that is less than actual observations have reported [the reported distances based on Hubble - Einstein cosmology].

That Hubble - Einstein cosmology problem is even more severe if the calculations of Table 5b are performed with no universal decay, as the Hubble - Einstein cosmology contends. The results for that case are presented in Table 5c below in which the greatest observable past distance is barely *8 Gyrs*, quite substantially less than reported observations. That is so even when a greatly more energetic case, *F = 3000*, is examined. For that extreme the energy is such that gravitation has not yet slowed that matter below, essentially, light speed; which means that its Doppler redshift would be nearly infinite, $z \approx \infty$.

```
     F-      ChangePoint    2nd Phase    Observable Past Distance    Relative!
   Factor   Time[Gyrs]    Constant-A    Back [Gyrs]    Then [Gyrs]   Abundance
     1        0.00471       16.59278       3.5544        10.4456        22.
     3        0.01413       16.64877       3.7343        10.2657        37.
    10        0.04713       16.71513       3.9972        10.0028        63.
    32        0.1518        16.79505       4.4214         9.5786        93.
   100        0.471         16.89360       5.1720         8.8280        90.
   316        1.489         17.01575       6.5490         7.4510        23.
  1000        4.710         17.16130       8.08125        5.91875        0.1
       ! = Estimate per Figure 1a, where Earth, F = 55, is of Abundance 100
```

```
  3000       14.13 [>Age]    n/a           8.15           5.85
```

*Table 5c*
*Table 5b Re-Calculated with No Universal Decay*
*and Added Extreme Case*

Clearly, the tenets of the Hubble - Einstein cosmology fail because they cannot conform to reality as it is already known. Returning to Universal Decay cosmology and the age of the universe being *Age = 30 Gyrs*, that age derives from what is needed to enable observation of redshifts on the order of *z = 10*, as presented in the next section below. It is an estimate because our instrumentation presently limits our ability to observe the past more than the theoretical limit does. Consequently new developments in instrumentation and observation may produce observed redshifts greater than `z = 10`, ones on the order of `z = 12` or more, and therefore require a corresponding increase in the estimated age of the universe.

The present value for the farthest back into the past that it is theoretically possible to observe regardless of the quality of our instrumentation is a little over `27 Gyrs` ago to the time `2 to 3 Gyrs` after the "Big Bang". The travel of the light source and of the observer's home and of the emitted light for that case of the most distant source theoretically observable, all from the time of the "Big Bang" to the present are as shown in Figure 6, below.

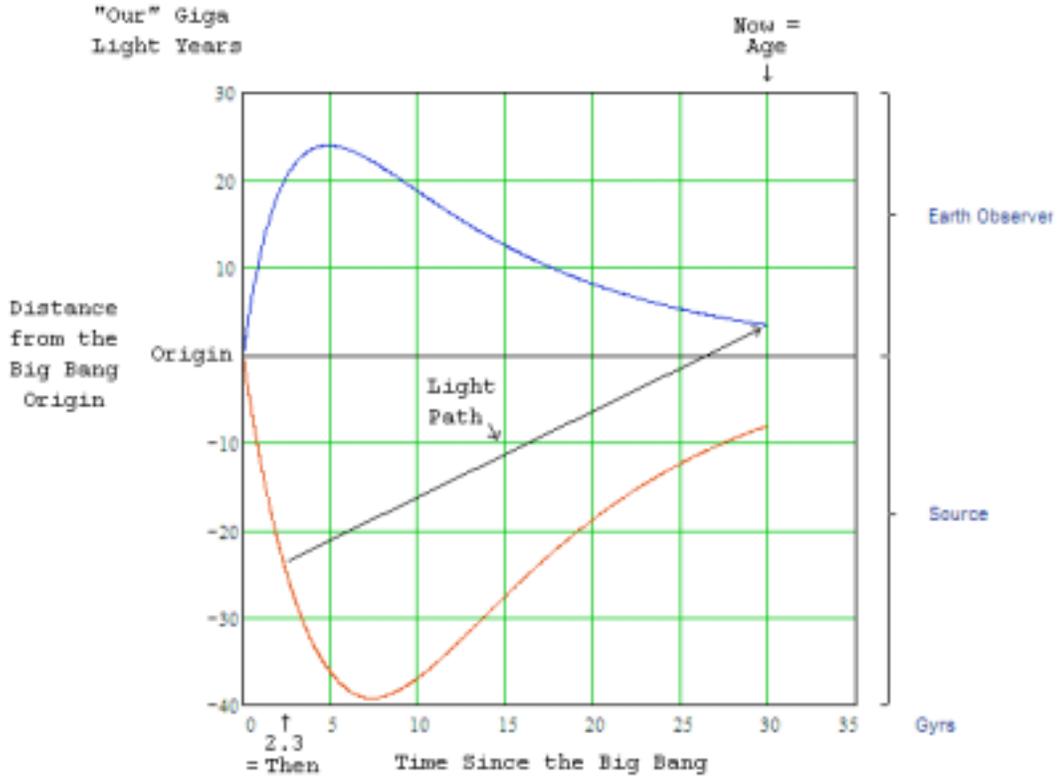

*Figure 6*
*The Most Distant into The Past Source [-] Observable by an Earth*
*Observer [+] the Two Diametrically Opposite Each Other Relative to the "Big Bang"*

### b. Redshifts: Universal Decay and Hubble Doppler

There are two principle causes of the redshifts that we observe: the effect of the universal decay and the Doppler shift due to astral objects' velocity away from us.

The universal decay redshift occurs because we observe ancient light traveling at the speed at which it was originally emitted, a speed significantly larger because less decayed than our present local speed of light. We observe the greater speed as a lengthening of all wavelengths in the light [with no change in frequencies]. The formulation for the universal decay redshift, $z_\tau$, of light that was emitted at time `t = T` after the "Big Bang" and is observed at time `t = now = age` is as follows.

(37) $z_\tau \equiv$ redshift due to the universal decay

$$= \frac{\lambda_{observed} - \lambda_{local}}{\lambda_{local}} = \frac{c(\text{time light emitted}) - c(\text{time now})}{c(\text{time now})}$$

*(37, continued)*

$$z_\tau = \frac{c(t=0)\cdot\varepsilon^{-[T/\tau]} - c(t=0)\cdot\varepsilon^{-[age/\tau]}}{c(t=0)\cdot\varepsilon^{-[age/\tau]}}$$

$$= \frac{\varepsilon^{-[T/\tau]}}{\varepsilon^{-[age/\tau]}} - 1$$

The formulation for the Doppler shift due to astral objects' velocity away from us, $z_D$, is as follows, per standard Hubble - Einstein cosmology.

*(38)* $z_D \equiv$ relativistic redshift due to the Doppler effect

$$= \frac{[1 + v/c]^{1/2}}{[1 - v/c]^{1/2}} - 1$$

The formulation for the universal decay redshift, equation *(37)*, is a function of time, not velocity. Equation *(38)* can be converted to expressing the Doppler redshift, $z_D$, in terms of time by using the velocity-as-a-function-of-time expressions for the motion of the astral body products of the "Big Bang" developed earlier above: equations *(24* and *27)*.

For the period from time $t = 0$ through $t =$ ChangePoint the velocity, equation *(24)*, is very nearly the then current decaying speed of light. The $v/c$ ratio is very nearly $1.0$ so that the redshifts, $z_D$, are very large, but are also essentially meaningless for any useful purpose. For the period from $t =$ ChangePoint onward the velocity expression is equation *(27)*, repeated below.

*(27)*
$$v2(t, Age, F) = c(t, Age) \cdot \frac{1}{\varepsilon^{A\cdot(t - ChangePoint)}} \quad \text{meters}/\text{sec}$$

where A and ChangePoint are as given in Table 5, above.

From equation *(27)* the $v/c$ ratio is:

*(39)*
$$v/c = \frac{1}{\varepsilon^{A\cdot(t - ChangePoint)}}$$

and by substituting that into equation *(38)* the expression for the Doppler redshift, $z_D$, is:

*(40)* $z_D \equiv$ relativistic redshift due to the Doppler effect

$$= \frac{[1 + v/c]^{1/2}}{[1 - v/c]^{1/2}} - 1$$

$$= \frac{\left[1 + \dfrac{1}{\varepsilon^{A\cdot(t - ChangePoint)}}\right]^{1/2}}{\left[1 - \dfrac{1}{\varepsilon^{A\cdot(t - ChangePoint)}}\right]^{1/2}} - 1$$

These two principle causes of redshifts are depicted independently in Figure 7, below.  Of course, the actual observed redshift is the sum of the two.  From the figure it is apparent that the Doppler-caused redshifts are quite minor until one is addressing light emitted only at times too early to be observable.

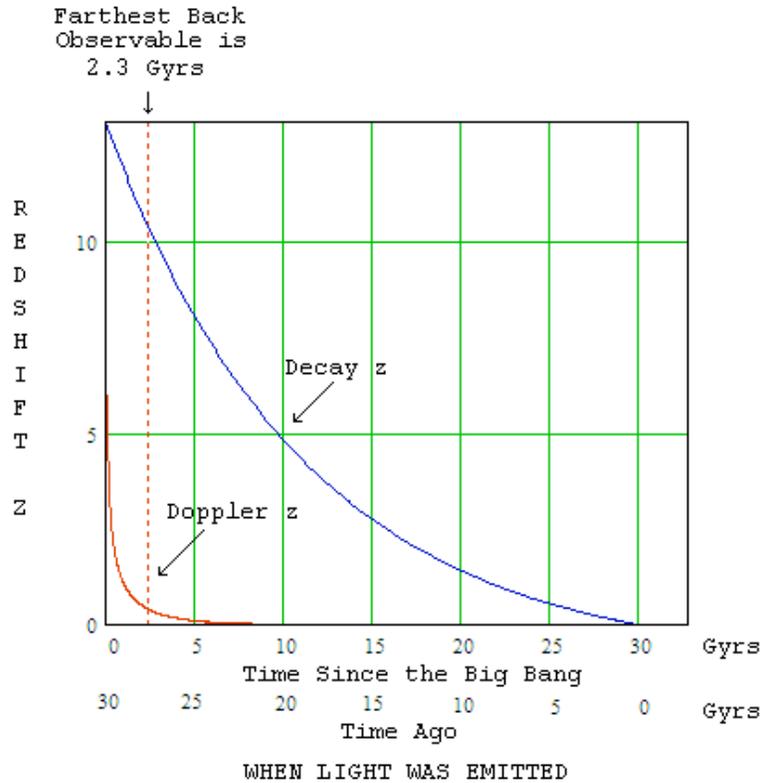

*Figure 7*
*Redshifts:  Caused by Universal Decay and by the Doppler Effect*

The figure also makes clear why the age of the universe must be on the order of *30 Gyrs*.  That amount of time is needed to include enough time constant periods, $\tau$, that is *11.3373 Gyrs*, each to yield a maximum observable redshift [at *then = 2.3 Gyrs*] of *z > 10* as in the figure.  A small number of redshifts at *z = 10* have been reported[5] with some indications of redshifts as high as *z = 12*.  Improved instrumentation and techniques may well result shortly in confirmed detection of redshifts at *z > 10*. The value *age = 30 Gyrs* is an, at present, conservative best estimate taking into account currently known observational data.

However, the more significant comparison of redshifts is to compare the decay-caused redshift to the redshift associated with the Hubble "Constant".  The development is as follows.

*(41)* Where: c ≡ "our" local light speed [the only light speed in Hubble
              - Einstein cosmology].
       v ≡ observed astral body's velocity away from us observers.
       d ≡ distance of observed astral body away from us observers.
       z ≡ observed redshift.
       $H_0$ ≡ Hubble "Constant".

   Then:   $H_0 = v/d$   *[km/sec/megaparsec]*         ["Hubble Law"]

           $v = c \cdot \dfrac{[z + 1]^2 - 1}{[z + 1]^2 + 1}$   *[km/sec]*     [equation (40) solved for "v"]

*(41, continued)*

$$d = \frac{v}{H_0} \quad [megaparsecs] \quad [\text{solving "Hubble Law" for "d"}]$$

$$= \frac{c}{H_0} \cdot \frac{[z+1]^2 - 1}{[z+1]^2 + 1} \quad [\text{substitute "v" from above}]$$

Three values for $H_0$ will be illustrated in recognition of the uncertainty of the correct value for it in Hubble - Einstein cosmology:

*(42)*  $H1 = 63 \ [km/sec/megaparsec] = 20.5 \cdot 10^6 \quad [meters/sec/G\text{-}Lt\text{-}Yr]$

$H2 = 75 \ [km/sec/megaparsec] = 24.4 \cdot 10^6 \quad [meters/sec/G\text{-}Lt\text{-}Yr]$

$H3 = 88 \ [km/sec/megaparsec] = 28.7 \cdot 10^6 \quad [meters/sec/G\text{-}Lt\text{-}Yr]$

and the resulting distances, *d*, produced by equation *(41)* using *c* in *[meters/sec]* and those values of $H_0$ in *[meters/sec/G-Lt-Yr]* are the same numerical value as the time from the present into the past in *Gyrs* [for the Hubble - Einstein value of *c*].

The variation of *z vs. t*, where *t = [time / distance into the past that the observed light was emitted]* per the above for the several values of $H_0$ is depicted in Figure 8, below, along with the corresponding Universal Decay variation of *z vs. t*.

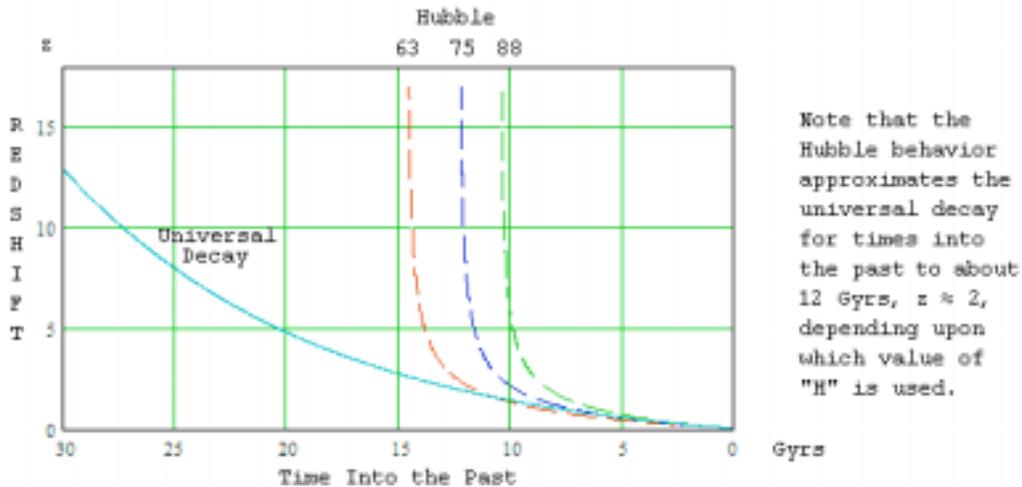

*Figure 8*
*Redshift Comparison of the Universal Decay and the Hubble - Einstein Concept*

The figure makes clear the reason for Hubble - Einstein cosmology estimates of the age of the universe being in the "mid-teens"; it is the asymptotic behavior of the relativistic Hubble redshift formulation. Unfortunately, the currently more favored value for the Hubble "Constant", $H_0 = 72$ fails to correspond well to the currently favored age of the universe of about *14 Gyrs*. It is also worth noting that the Type Ia Supernovae observations that are the basis for the contended "dark energy" were at redshifts in the range *z = 0.4 - 0.8*, astral objects at up to about *5 Gyrs* into the past.

### 3 - *The Fate of the Universe*

Although the initial velocities of the "Big Bang" product particles were greater than the there / then escape velocity, as shown earlier above, their velocities now are all well under their escape velocity. We are used to escape velocity being escape velocity -- a simple yes or no proposition. The reason that that is not the case for the overall universe is the effect of the speed of light limitation.

The escape process is the conversion of kinetic energy into gravitational potential energy. If the initial kinetic energy is greater than the maximum possible gravitational potential energy then there will be escape. In the case of a rocket leaving Earth that process is accompanied by the rocket's velocity taking it farther enough away from the Earth that the gravitational effect is reduced in "proper" relation to the process. But the "Big Bang" product particles were not permitted to so travel, their actual velocity being limited to just under light speed as compared to the much larger theoretical non-relativistic velocity at which they would have had to have traveled outward for the accrued distance to correspondingly reduce the gravitational effect in "proper" relation to the process.

The actual velocities and the related escape velocities now at time $t = Age$ for the same distribution of initial "Big Bang" product energies analyzed earlier above is given in Table 9, below. As in the earlier analysis of the initial escape velocity, the present analysis is non-relativistic, using velocities greater than the speed of light rather than letting mass relativistically increase.

```
For:  Universe Age = 30 Gyrs, which means that:
      Initial Light Speed = 4.226,895,62·10^9 m/s
      Initial Gravitation Constant, G = 1.870,24·10^-7 m^3/kg-s^2
 At Age = 30 Gyrs:

    F-            Outward from the Origin of the "Big Bang"*
  Factor      Velocity[m/s]    Distance[G-Lt-Yrs]   Escape Velocity[m/s]

    1         0.00003814·10^8       2.157              7.418·10^8
    3         0.0001505 ·10^8       2.400              7.032·10^8
   10         0.0006230 ·10^8       2.727              6.597·10^8
   32         0.002731  ·10^8       3.205              6.085·10^8
   55 Earth   0.003700  ·10^8       3.374              5.931·10^8
  100         0.01287   ·10^8       3.980              5.461·10^8
  316         0.06673   ·10^8       5.388              4.693·10^8
 1000         0.3974    ·10^8       8.034              3.844·10^8

  *=Decayed to Age
```

*Table 9*
*Actual Velocities vs. Escape Velocities Now, at $t = Age$*

One must immediately conclude that the entire material universe is ultimately destined to collapse back toward the location of its origin, just as a ball tossed straight up from the Earth's surface ultimately returns to its starting point. However, the case of the universe is more complicated than that of the simple ball and there are also two different considerations for the case of the universe: its matter and its radiation.

### a. *The Fate of the Universe's Radiation*

The fate of the radiation emitted from sources [primarily astral sources] throughout the universe is very different from the fate of the universe's matter. Most of the universe's radiation continues propagating outward forever, reduced in concentration inversely as the square of the distance from its source, and carrying outward in itself a significant amount of the universe's energy, which energy becomes essentially lost to the remainder of the universe, the universe's matter. That comes about as follows.

### (1) Gravitational Redshift and Light Escape

When a particle of mass $m$ climbs in a gravitational field its speed is reduced by the gravitation, which speed reduction reduces its kinetic energy, $\tfrac{1}{2} \cdot m \cdot v^2$. Conservation is maintained by the kinetic energy loss being replaced by gravitational potential energy increase.

A photon of frequency $f$ has kinetic mass, $m_{ph}$, [even though it has no rest mass].

(43)  $m_{ph} = \text{energy}/c^2$

   $= h \cdot f / c^2$

As light, with its kinetic mass, $m_{ph}$, climbs in a gravitational field, instead of its speed being reduced its frequency is shifted lower [toward the red]. The photon cannot slow down [to correspondingly reduce its kinetic energy as a particle of matter would] because it is constrained by its nature to only travel at light speed, $c$. Instead the photon frequency, $f$, decreases, which reduces its energy, $h \cdot f$, its energy of motion that corresponds to kinetic energy.

Then, for a photon to be able to escape from a gravitational field in a manner analogous to escape for a particle of mass, the photon energy, $h \cdot f$, must at least just exceed the depth of the gravitational potential energy pit, $G \cdot M \cdot m_{ph}/R$, that it experiences at the location where the photon is emitted. On that basis the calculation for photon escape would be that the photon frequency must be at least such that

(44)  $h \cdot f_{minimum} > G \cdot M \cdot m_{ph}/R$

however, the photon mass, $m_{ph}$, depends on $f$ per equation (43) so that a directly solvable relationship cannot be obtained on that basis; photon escape is independent of photon frequency.

### (2) The "Schwarzschild Radius" and Escape

Astrophysicists treat a quantity called the "Schwarzschild Radius". The line of thought is that the depth of a gravitational potential energy pit from which a particle must climb in order to escape is $G \cdot M \cdot m / R$ where $G$ is the gravitation constant, $M$ is the gravitating mass, $m$ is the mass of the particle attempting escape, and $R$ is the distance from the center of the gravitating mass at which the particle must begin its attempt. To escape, the particle's kinetic energy, $\tfrac{1}{2} \cdot m \cdot v^2$, must just exceed that potential energy so that, as presented earlier, the escape velocity is $v_{esc} = [2 \cdot G \cdot M / R]^{\tfrac{1}{2}}$.

From that formulation, as $R$ decreases the required velocity, $v$ increases. Therefore one can calculate a radius, $R_S$, the "Schwarzschild Radius", for any particular gravitating body mass, $M$, such that the required escape velocity, $v_{esc}$, is the speed of light, $c$, as follows.

(45) For Light, a Photon of Mass $m_{ph}$:

   Photon Energy = Gravitational Potential Energy

   $m_{ph} \cdot c^2 = G \cdot M \cdot m_{ph}/R = G \cdot M \cdot m_{ph}/R_S$

   $R_S = G \cdot M / c^2$

(46) For a Particle of Mass "m":

   Kinetic Energy = Gravitational Potential Energy

   $\tfrac{1}{2} \cdot m \cdot v^2 = G \cdot M \cdot m / R$

   $m \cdot c^2 = G \cdot M \cdot m / R_S$   [Because KE = TotalE − RestE, then as v → c
                      TotalE >> RestE and KE → TotalE not ½·TotalE
                      all because of the relativistic mass increase]

   $R_S = G \cdot M / c^2$               [Solve for $R_S$]

                      [The usual presentation, that ignores the
                       effect as in the above note, is $R_S = \underline{2} \cdot G \cdot M / c^2$]

No matter can travel at light speed, therefore matter located at or nearer to the center of the gravitating mass than $R_S$ cannot ever escape. For radiation escape is independent of the frequency and depends only on the distance, $R_S$.

For the value of $R_S$ for the universe at the instant of the "Big Bang":

· from equation *(19)* G was $G(0) = 1.870 \cdot 10^{-7}$ m³/kg-s²,

· from equation *(16)* M was $m_{Universe} = 3 \cdot 10^{49}$ kg, and

· from equation *(6)* c was $c(0) = 4.226 \cdot 10^9$ meters/sec.

Then, the value of $R_S$ for the universe at the instant of the "Big Bang" was *$3.14 \cdot 10^{23}$ meters* which is *[0.033 G-Lt-Yrs]* and at that instant the actual distance from the center of the "Big Bang" was much less, $d_0 = 4.0 \cdot 10^7$ *meters*. Therefore, at that time, *t = 0*, no matter nor light could escape from the "Big Bang" as already demonstrated and summarized for matter in Table 9, above. The inability of matter to escape did not change thereafter.

However, in its rapid initial expansion at a speed of very nearly *c(0)*, after time approximately *[$R_S \div c(0)$]*, that is the first two to three million years, the light-source matter of the universe had moved out from the origin to beyond the "Schwarzschild Radius" and whatever radiation was emitted thereafter was free to travel outward forever.

### b. *The Fate of the Universe's Matter*

The universe's matter, however, was already embedded in the impossibility of escape and it only remains to investigate its fate.

To this point the material presented has consisted of analytical deductions and reasonable estimates based on fundamentals of physics, the available data, and the tenets of the theories involved. Now, with regard to the fate of the universe's matter, some of what is presented must be limited to "educated" speculation as to the implied future while some still remains reasonable analytical deductions.

Clearly the large range of the present velocities of the universe's matter and of its varied present distances outward from the origin per Table 9 means that the universe's matter's gradual slowing - direction reversal - inward collapse will result in a wide range of arrival times at the origin of the original expansion of the universe's various portions. [That as juxtaposed to the concept of a universe all together collapsing and then re-exploding outward in a succession of "big bangs" as has been hypothesized in the not too distant past.] There are, then, several possibilities to be considered.

- Matter arriving at the initial origin crashing into like kind arriving matter.

    Recently there have been analyses of what happens when a large asteroid crashes into the Earth, the energies involved and the resulting destruction being immense. One must [speculatively] increase those energies and their results many orders of magnitude to conceive of what would happen at the collision of two planetary bodies, two suns, or two galaxies.

    However, the collision would be kinetic and produce great heat, breakdown into particles, and great kinetic energy of those product particles. It would not be as a nuclear fission nor fusion explosion, that is it most likely would not involve a major conversion of matter to energy.

- Matter arriving at the initial origin encountering there nothing but empty space.

    Unlike the case of the ball tossed upward from the Earth's surface in which case the Earth is still there when the ball falls back down, it would seem that there is now likely nothing but empty space at the location of the initial origin. A portion of the universe's matter arriving there unopposed would be traveling at

high speed [most likely the same [then outward but now inward] speed as was imparted to it in the original "Big Bang" [but as reduced by the Universal Decay of the speed of light]. That body of matter would pass on through and proceed outward again in its own "personal" replay of its earlier role.

Except, that is, that the first time the gravitating matter of the universe was initially all concentrated at the origin whereas the second time that matter is scattered over a large universe volume. The gravitational conditions would be different for the second pass and the escape velocity would also be different. One can only [speculatively] imagine various scenarios for the further travel of that portion of the universe's matter and its peers / partners.

- Matter arriving at the initial origin and there encountering anti-matter.

There are two alternative hypotheses that can be considered with regard to anti-matter creation in the "Big Bang":

· Anti-matter was created, but in a lesser amount than ordinary-matter, and quite shortly thereafter all of the anti-matter mutually annihilated with an equal amount of ordinary-matter leaving essentially no remaining anti-matter and a small remaining amount of ordinary-matter, which is the matter of our universe. In this hypothesis there is no, or negligible anti-matter in today's universe.

This alternative voids the "matter arriving encountering anti-matter" possibility.

· Matter and anti-matter were created in equal, "mirror" amounts and, while most of it promptly mutually annihilated, small equal amounts of each participated in the outward expansion quickly enough to survive. Thus our universe has matter portions [galaxies, galaxy groups, etc.] and similar anti-matter portions and their continued separation in space largely preserves their continued independent existence.

In this hypothesis matter arriving at the initial origin could encounter anti-matter, which would result in a mutual annihilation. Unlike the kinetic collision case, the result would be an immense amount of energy radiated as gamma rays. Such events involving significant bodies of matter could be the cause of the extremely high energy gamma ray bursts that have been observed but remain un-accounted-for.

· With regard to the two alternative hypotheses:

The first, the present universe essentially lacking anti-matter, would appear to require anti-matter to differ from ordinary-matter other than by being a perfect "mirror". That would appear to conflict with a symmetrical "Big Bang" as required for conservation. Nevertheless, research seeking to discover a discriminating non-symmetrical difference between matter and anti-matter is being conducted without definitive results so far.

The second is what would be expected for a perfectly symmetrical "Big Bang" consistent with conservation. That alternative is pursued in the remainder of this analysis.

The behavior of anti-matter is such that there is no way to discriminate whether a distant astral source is matter or anti-matter: the gravitation is the same; the light emitted is the same.

### *c. The Ultimate End of the Universe.*

#### *(1) The Universal Decay Will Continue*

The universe will continue shrinking to beyond the point of extremely minute, all to no noticeable effect on its internal functioning no matter how small it becomes relative to the size that it is now or originally was.

If one looks back one million years ago, lengths then were greater than the corresponding lengths today by a factor of `1.0009`. Clearly the universal decay has little significance in day to day life. In fact, its only significance is for astronomers, because only they can look back into the past far enough to see the effects of the extremely slow decay.

Everything decays proportionately. The ratio at any time, now or in the past or in the future, of the size of things relative to things does not change at all. There is no fixed objective reference by which one could appreciate or notice the decay other than those accessible only to astronomy. Everything is shrinking, but to no noticeable effect. Whatever happens to be left of the universe some inconceivable number of aeons from now will be so extremely minute compared to the size of things in today's universe as to seem to us as nothing. Yet it will operate, function, behave according to the same rules as our universe now, as if it had not decayed at all [again except astronomically], but subject to the events below.

#### *(2) The Universe's Matter Will Gradually Completely Obliterate*

Whatever time it takes, eventually all of the universe's matter will be obliterated in mutual annihilations. The process will be a kind of universe "Russian Roulette", annihilations depending randomly on the simultaneous arrival of matter and anti-matter portions of the universe at the location of the initial origin. Such annihilations will extend only to the extent of arriving masses being equal; the un-annihilated surplus of the greater being hurled outward again for another excursion and later chance of annihilation upon its return.

#### *(3) The Universe's Radiation and Energy Will Be Dispersed in Endless Space*

All of the radiation and energy of the matter annihilations along with all of the astral and other radiation and energy from the beginning on [including radiation absorbed and subsequently re-radiated] will disperse outward in space, gradually reddening and so reduced by inverse square dispersion as to eventually amount to essentially nothing.

#### *(4) Nothing to Nothing …*

In the same way as for we humans when our span of life ends it is said, "Ashes to ashes and dust to dust ....", so for the universe it can be said, "It came from nothing and eventually passes on to nothing, to that from which it came".